%% file: main.tex
\documentclass{article}

\usepackage{microtype}
\usepackage{graphicx}
\usepackage{subfigure}
\usepackage{booktabs} % for professional tables
\usepackage{multirow}
\usepackage{soul}

\include{todonotes}

\usepackage{hyperref}

\usepackage{xspace}
\newcommand{\tdd}{TDD-Bench-Verified\xspace}
\newcommand{\solx}{Otter\xspace}
\newcommand{\soly}{Otter++\xspace}

\newcommand{\etal}{\emph{et al.}\xspace}

\usepackage[accepted]{icml2025}

\usepackage{amsmath}
\usepackage{amssymb}
\usepackage{mathtools}
\usepackage{amsthm}

\usepackage[capitalize,noabbrev]{cleveref}

\theoremstyle{plain}

\theoremstyle{definition}

\theoremstyle{remark}

\usepackage[textsize=tiny]{todonotes}
\usepackage{alltt}
\usepackage{enumitem}
\icmltitlerunning{Otter: Generating Tests from Issues to Validate SWE Patches}

\begin{document}

\twocolumn[
\icmltitle{Otter: Generating Tests from Issues to Validate SWE Patches}

% It is OKAY to include author information, even for blind
% submissions: the style file will automatically remove it for you
% unless you've provided the [accepted] option to the icml2025
% package.

% List of affiliations: The first argument should be a (short)
% identifier you will use later to specify author affiliations
% Academic affiliations should list Department, University, City, Region, Country
% Industry affiliations should list Company, City, Region, Country

% You can specify symbols, otherwise they are numbered in order.
% Ideally, you should not use this facility. Affiliations will be numbered
% in order of appearance and this is the preferred way.
\icmlsetsymbol{equal}{*}

\begin{icmlauthorlist}
    \icmlauthor{Toufique Ahmed}{yyy}
    \icmlauthor{Jatin Ganhotra}{yyy}
    \icmlauthor{Rangeet Pan}{yyy}
    \icmlauthor{Avraham Shinnar}{yyy}
    \icmlauthor{Saurabh Sinha}{yyy}
    \icmlauthor{Martin Hirzel}{yyy}
    \end{icmlauthorlist}
    
    \icmlaffiliation{yyy}{IBM Research, Yorktown Heights, New York, USA}

    \icmlcorrespondingauthor{Toufique Ahmed}{tfahmed@ibm.com}
    \icmlcorrespondingauthor{Martin Hirzel}{hirzel@us.ibm.com}

%\icmlaffiliation{comp}{Company Name, Location, Country}
%\icmlaffiliation{sch}{School of ZZZ, Institute of WWW, Location, Country}

%\icmlcorrespondingauthor{Firstname1 Lastname1}{first1.last1@xxx.edu}
%\icmlcorrespondingauthor{Firstname2 Lastname2}{first2.last2@www.uk}

% You may provide any keywords that you
% find helpful for describing your paper; these are used to populate
% the "keywords" metadata in the PDF but will not be shown in the document
\icmlkeywords{LLMs, SWE Patches, Reproduction Tests}

\vskip 0.3in
]

% this must go after the closing bracket ] following \twocolumn[ ...

% This command actually creates the footnote in the first column
% listing the affiliations and the copyright notice.
% The command takes one argument, which is text to display at the start of the footnote.
% The \icmlEqualContribution command is standard text for equal contribution.
% Remove it (just {}) if you do not need this facility.

\printAffiliationsAndNotice{}  % leave blank if no need to mention equal contribution
%\printAffiliationsAndNotice{\icmlEqualContribution} % otherwise use the standard text.

\begin{abstract}
While there has been plenty of work on generating tests from existing
code, there has been limited work on generating tests from issues.
A correct test must validate the code patch that resolves the issue.
This paper focuses on the scenario where that code patch does not yet exist.
Doing so supports two major use-cases.
First, it supports TDD (test-driven development), the discipline of
``test first, write code later'' that has well-documented benefits for
human software engineers.
Second, it also validates SWE (software engineering) agents, which generate code patches for resolving issues.
This paper introduces \tdd, a benchmark for generating tests from
issues, and \solx, an LLM-based solution for this task.
\solx augments LLMs with rule-based analysis to check and repair their
outputs, and introduces a novel self-reflective action planner.
Experiments show \solx outperforming state-of-the-art systems for
generating tests from issues, in addition to enhancing
systems that generate patches from issues.
We hope that Otter helps make developers more productive
at resolving issues and leads to more robust, well-tested code.
\end{abstract}

\input{intro}
\input{problem}
\input{relatedwork}
\input{method}
\input{benchmark}
\input{result}

\input{limitation}

\input{conclusion}

\section*{Impact Statement}

This paper presents work whose goal is to advance the field of software testing and program repair. Developers spend a significant amount of time resolving bugs and testing them. We believe this work will significantly improve the developers' experience in their day-to-day life.  We envision two ways of integrating Otter into existing developer workflows. First, Otter can run on a new issue to propose a test, which the stakeholders can use to clarify requirements for the desired behavior after issue resolution and, following that, a developer can use the test for test-driven development to resolve the issue. Second, Otter can be paired with a patch-generation solution (a ``SWE agent'') to create a PR that includes both a patch and a test.

%Authors are \textbf{required} to include a statement of the potential 
%broader impact of their work, including its ethical aspects and future 
%societal consequences. This statement should be in an unnumbered 
%section at the end of the paper (co-located with Acknowledgements -- 
%the two may appear in either order, but both must be before References), 
%and does not count toward the paper page limit. In many cases, where 
%the ethical impacts and expected societal implications are those that 
%are well established when advancing the field of Machine Learning, 
%substantial discussion is not required, and a simple statement such 
%as the following will suffice:
%
%``This paper presents work whose goal is to advance the field of 
%Machine Learning. There are many potential societal consequences 
%of our work, none which we feel must be specifically highlighted here.''
%
%The above statement can be used verbatim in such cases, but we 
%encourage authors to think about whether there is content which does 
%warrant further discussion, as this statement will be apparent if the 
%paper is later flagged for ethics review.
%

% In the unusual situation where you want a paper to appear in the
% references without citing it in the main text, use \nocite

\bibliography{bibfile}
\bibliographystyle{icml2025}

%%%%%%%%%%%%%%%%%%%%%%%%%%%%%%%%%%%%%%%%%%%%%%%%%%%%%%%%%%%%%%%%%%%%%%%%%%%%%%%
%%%%%%%%%%%%%%%%%%%%%%%%%%%%%%%%%%%%%%%%%%%%%%%%%%%%%%%%%%%%%%%%%%%%%%%%%%%%%%%
% APPENDIX
%%%%%%%%%%%%%%%%%%%%%%%%%%%%%%%%%%%%%%%%%%%%%%%%%%%%%%%%%%%%%%%%%%%%%%%%%%%%%%%
%%%%%%%%%%%%%%%%%%%%%%%%%%%%%%%%%%%%%%%%%%%%%%%%%%%%%%%%%%%%%%%%%%%%%%%%%%%%%%%
\newpage
\appendix
\onecolumn
\section{Different attributes of the \tdd Instances \& Sample Test Patch}

\begin{figure}[h]
    \centering
    \includegraphics[trim=0 0cm 0 0cm, width=0.6\columnwidth]{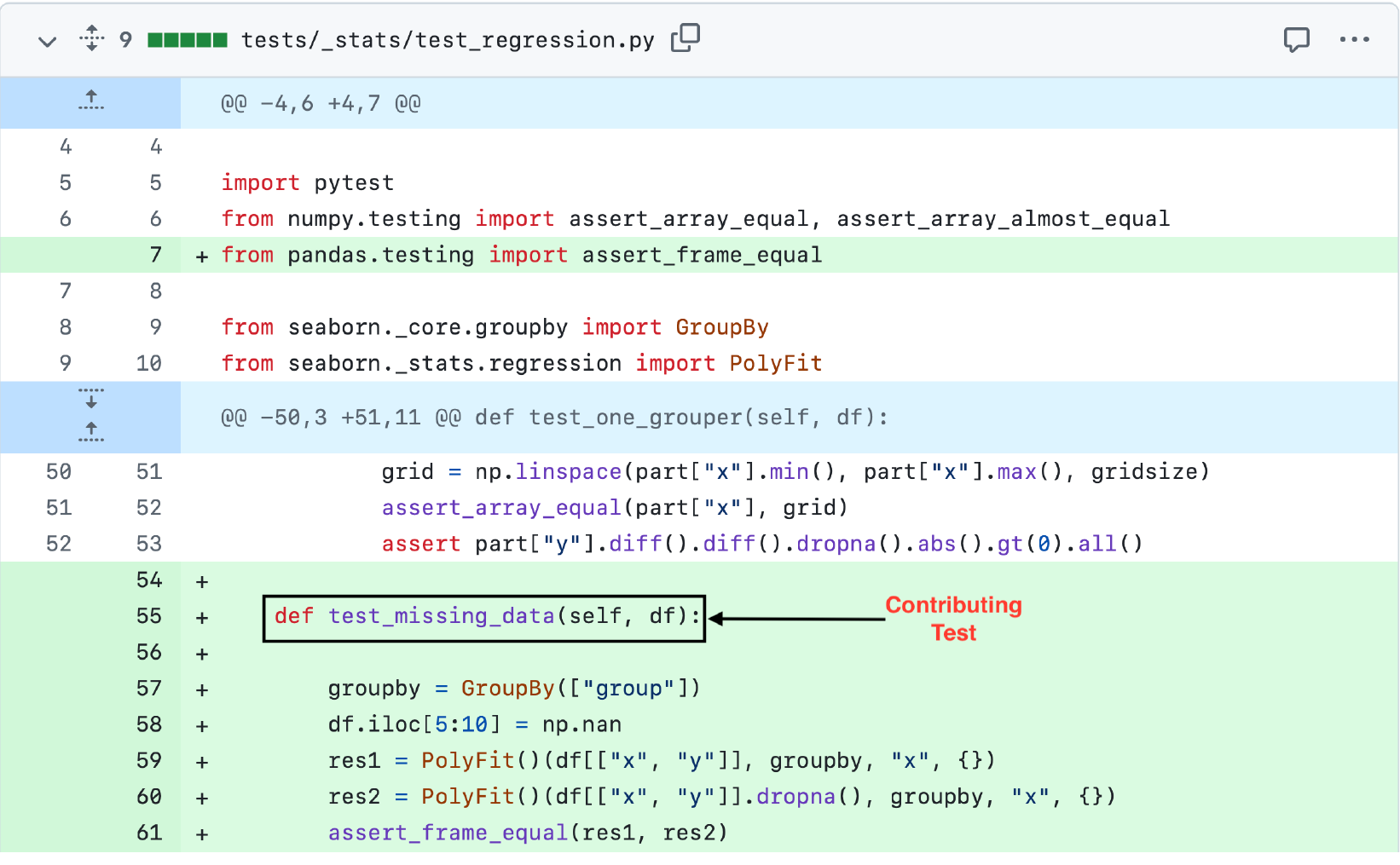}
    \caption{Example test patch with one contributed test. Although the test file name \texttt{\small test\_regression.py} and test name \texttt{\small test\_missing\_data} are available in this diff, the class \texttt{\small TestPolyFit} enclosing \texttt{\small test\_missing\_data} is missing. By applying the test patch to the base commit and parsing the file, we retrieve \texttt{\small TestPolyFit}, which is required to run \texttt{\small test\_missing\_data}.}
    \label{fig:test_patch}
    \vspace{-15pt}
\end{figure}

\cref{tbl:tdd-stat} presents different attributes of the TDD instances to give some idea about the nature of the issues we are dealing with in \tdd. We also present a sample test patch in ~\cref{fig:test_patch}.

\begin{table*}[h]
\centering
\caption{Different attributes of the \tdd instances.}
\resizebox{.9\textwidth}{!}{%
\renewcommand{\arraystretch}{1.2}% Tighter
\begin{tabular}{lrrrrrrr}
\toprule
\multicolumn{1}{c}{\multirow{2}{*}{Project}} & \multicolumn{1}{c}{\multirow{2}{*}{\# of Instances}} & \multicolumn{1}{c}{\multirow{2}{*}{\begin{tabular}[c]{@{}c@{}}Fraction of \\ Dataset (in \%)\end{tabular}}} & \multicolumn{1}{c}{\multirow{2}{*}{\# of Files}} & \multicolumn{1}{c}{\multirow{2}{*}{\# of Test Files}} & \multicolumn{2}{c}{\begin{tabular}[c]{@{}c@{}}Average \# of Lines \\ Deleted and Added\end{tabular}} & \multicolumn{1}{c}{\multirow{2}{*}{\begin{tabular}[c]{@{}c@{}}Average Word Count\\ in Issue Description\end{tabular}}} \\ \cmidrule{6-7}
\multicolumn{1}{c}{}                          & \multicolumn{1}{c}{}                                & \multicolumn{1}{c}{}                                                                                        & \multicolumn{1}{c}{}                             & \multicolumn{1}{c}{}                                  & \multicolumn{1}{c}{On Code}                         & \multicolumn{1}{c}{On Tests}                        & \multicolumn{1}{c}{}                                                                                                   \\ \midrule
Astropy                                       & 18                                                  & 4.0                                                                                                        & 1,990                                             & 351                                                   & 11.9                                                & 28.7                                                & 304.5                                                                                                                  \\
Django                                        & 212                                                 & 47.2                                                                                                       & 6,863                                             & 810                                                   & 12.0                                                  & 24.7                                                & 145.6                                                                                                                  \\
Flask                                         & 1                                                   & 0.2                                                                                                        & 275                                              & 27                                                    & 3.0                                                   & 5.0                                                  & 35.0                                                                                                                     \\
Matplotlib                                    & 32                                                  & 7.1                                                                                                        & 4,656                                             & 102                                                   & 9.3                                                 & 20.0                                                  & 260.5                                                                                                                  \\
Pylint                                        & 10                                                  & 2.2                                                                                                       & 3,833                                             & 51                                                    & 24.7                                                & 33.8                                                & 347.1                                                                                                                  \\
Pytest                                        & 16                                                  & 3.6                                                                                                       & 639                                              & 114                                                   & 24.6                                                & 53.5                                                & 250.1                                                                                                                  \\
Requests                                      & 5                                                   & 1.1                                                                                                       & 155                                              & 9                                                     & 3.6                                                 & 6.6                                                 & 85.2                                                                                                                   \\
Scikit-learn                                  & 25                                                  & 5.6                                                                                                        & 1,772                                             & 242                                                   & 11.8                                                & 17.1                                                & 297.6                                                                                                                  \\
Seaborn                                       & 2                                                   & 0.5                                                                                                        & 353                                              & 34                                                    & 13.5                                                & 18.5                                                & 182.5                                                                                                                  \\
Sphinx                                        & 41                                                  & 9.1                                                                                                        & 1,917                                             & 137                                                   & 17.5                                                & 26.1                                                & 186.2                                                                                                                  \\
Sympy                                         & 67                                                  & 14.9                                                                                                       & 2,050                                             & 617                                                   & 12.1                                                & 11.9                                                & 114.2                                                                                                                  \\
Xarray                                        & 20                                                  & 4.5                                                                                                       & 394                                              & 67                                                    & 17.1                                                & 24.3                                                & 301.0                                                                                                                    \\ \midrule
Overall                                       & 449                                                 & 100.0                                                                                                         & 24,897                                            & 2,561                                                  & 13.2                                                & 23.3                                                & 182.0     \\ \bottomrule                               \multicolumn{8}{l} {*File counts are based on the main branches of the project (cloned on October 29, 2024).}
                                                                               
\end{tabular}
}
\label{tbl:tdd-stat}
\end{table*}

%\newpage

\section{Action Counts in Self-Reflective Action Planner Phase}
\label{appendix_action_counts}

\cref{tbl:action_count} presents the stats on the different types of actions proposed by the model in the self-reflective action planner phase.

\begin{table}[h]
\centering
\caption{Stats on action count.}
\vskip 0.05in
\resizebox{.3\columnwidth}{!}{%
\renewcommand{\arraystretch}{1.2}% Tighter
\begin{tabular}{llrrr}
\toprule
\multicolumn{1}{c}{Model}      & \multicolumn{1}{c}{Action} & \multicolumn{1}{c}{Average} & \multicolumn{1}{c}{Max} & \multicolumn{1}{c}{Min} \\ \midrule
\multirow{3}{*}{GPT-4o}        & Read                       & 5.4                         & 21                      & 3                       \\
                               & Write                      & 1.2                         & 6                       & 0                       \\
                               & Modify                     & 0.5                         & 5                       & 0                       \\ \midrule
\multirow{3}{*}{Mistral-large} & Read                       & 6.5                         & 20                      & 2                       \\
                               & Write                      & 1.2                         & 5                       & 0                       \\
                               & Modify                     & 0.5                         & 5                       & 0                      \\ \bottomrule
\end{tabular}}
\label{tbl:action_count}
%\vspace{-10pt}
\end{table}

\section{Scaling of \soly}
\label{scaling}
\soly uses heterogeneous prompting and execution logs to select the best solution. Therefore, we cannot significantly increase the number of samples. \soly scaled well with samples up to 5, giving 0.5\%-5.6\% improvement (see~\cref{tbl:scaling}).

\begin{table}[h]
    \centering
    \caption{Scaling of \soly}
    \vskip 0.05in
    \resizebox{.3\columnwidth}{!}{%
    \renewcommand{\arraystretch}{1.2}% Tighter
    \begin{tabular}{rrr}
        \toprule
        \multicolumn{1}{c}{\# of Candidate} & \multicolumn{1}{c}{\# of Fail-to-pass} & \multicolumn{1}{c}{In \%} \\ \midrule
        1                                   & 141                                    & 31.4                      \\
        2                                   & 154                                    & 34.3                      \\
        3                                   & 157                                    & 35.0                      \\
        4                                   & 164                                    & 36.5                      \\
        5                                   & 166                                    & 37.0      \\ \bottomrule                
        \end{tabular}}
    \label{tbl:scaling}
    %\vspace{-10pt}
    \end{table}

\section{Heterogenous Prompts vs Temperature}
\label{heterogeneous_prompt}

In~\cref{tbl:temperature}, the Top-1 column shows ``fail-to-pass @ 1'' based on our actual ranker, whereas the Oracle column shows ``fail-to-pass @ N'', i.e., the result if we had a perfect ranker. Heterogeneous prompts yield better results than high-temperature samples in both the Top-1 column and the Oracle column. This is evidence that the difference indeed comes from the diversity of heterogeneous prompts and is not ranker specific. The difference between the Top-1 and the Oracle column is almost the same in the first three rows (i.e., all settings with 5 samples). That said, there is still room for improvement in the ranker, as illustrated by the last row. In the last row, an oracle that could reliably choose among all 10 samples would perform amazingly well, motivating further work on better rankers.

\begin{table}[t]
    \centering
    \caption{Performance of Our Test Generation Approach in Different Settings.}
    %(not \tdd).}
    \vskip 0.05in
    \resizebox{.9\columnwidth}{!}{%
    \renewcommand{\arraystretch}{1.2}% Tighter
    \begin{tabular}{llrrrr}
        \toprule
      \multicolumn{1}{c}{Setting}               & \multicolumn{1}{c}{Individual Performance}      & \multicolumn{1}{c}{Max} & \multicolumn{1}{c}{Average} & \multicolumn{1}{c}{Top-1} & \multicolumn{1}{c}{Oracle} \\ \midrule
      5 heterogeneous prompts (Otter++)         & 141, 110,115, 107, 110                          & 141                     & 116.6                       & 166                       & 197                        \\
      5 high-temperature samples (temp=1.0)     & 111, 118, 126, 121, 113                         & 126                     & 117.8                       & 146                       & 173                        \\
      5 = 1 greedy + 4 high-temperature samples & 141, 111, 118, 126, 121                         & 141                     & 123.4                       & 152                       & 183                        \\
      10 = 5 heterogeneous + 5 high-temperature & 141, 110,115, 107, 110, 111, 118, 126, 121, 113 & 141                     & 117.2                       & 168                       & 218        \\ \bottomrule               
      \end{tabular}
    }
    \label{tbl:temperature}
    \vspace{-10pt}
    \end{table}

    \section{Impact of hallucination replacer.}
    \label{halucination}

    The localizer makes two LLM calls during which hallucinations can occur. In the first call, we drop hallucinated file names. In the second call, we replace hallucinated file names with existing ones to ensure our pipeline functions correctly.
    ~\cref{tbl:hallunication} show that our proposed technique helps Otter achieve better performance. For example, if we didn’t replace the hallucinated file name in the 2nd call of test localization (the last row), we would have lost 5 samples because our pipeline would have exited with an error. Replacing the hallucinated name generated 4 fail-to-pass tests for those samples.

    \begin{table}[t]
      \centering
      \caption{Impact of hallucination replacer on GPT-4o generated tests.}
      %(not \tdd).}
      \vskip 0.05in
      \resizebox{.5\columnwidth}{!}{%
      \renewcommand{\arraystretch}{1.2}% Tighter
      \begin{tabular}{llrrr}
        \toprule
        \multicolumn{1}{c}{Focal/Test} & \multicolumn{1}{c}{LLM Call} & \multicolumn{1}{c}{\begin{tabular}[c]{@{}c@{}}\#of Halluciantion \\ Happen\end{tabular}} & \multicolumn{1}{c}{\#of fail-to-pass} & \multicolumn{1}{c}{\begin{tabular}[c]{@{}c@{}}\#of fail-to-pass \\ rate\end{tabular}} \\ \midrule
        \multirow{2}{*}{Focal}         & LLM Call 1                   & 19                                                                                       & 8                                     & 42.1                                                                                  \\
                                       & LLM Call 2                   & 5                                                                                        & 3                                     & 60.0                                                                                    \\ \midrule
        \multirow{2}{*}{Test}          & LLM Call 1                   & 39                                                                                       & 22                                    & 56.4                                                                                  \\
                                       & LLM Call 2                   & 5                                                                                        & 4                                     & 80.0      \\ \bottomrule                                                                              
        \end{tabular}
      }
      \label{tbl:hallunication}
      \vspace{-10pt}
      \end{table}

\section{Data Contamination}

\begin{figure}[h!]
    \centering
    \begin{minipage}[b]{0.8\textwidth}
    \centering
    \includegraphics[width=.7\columnwidth]{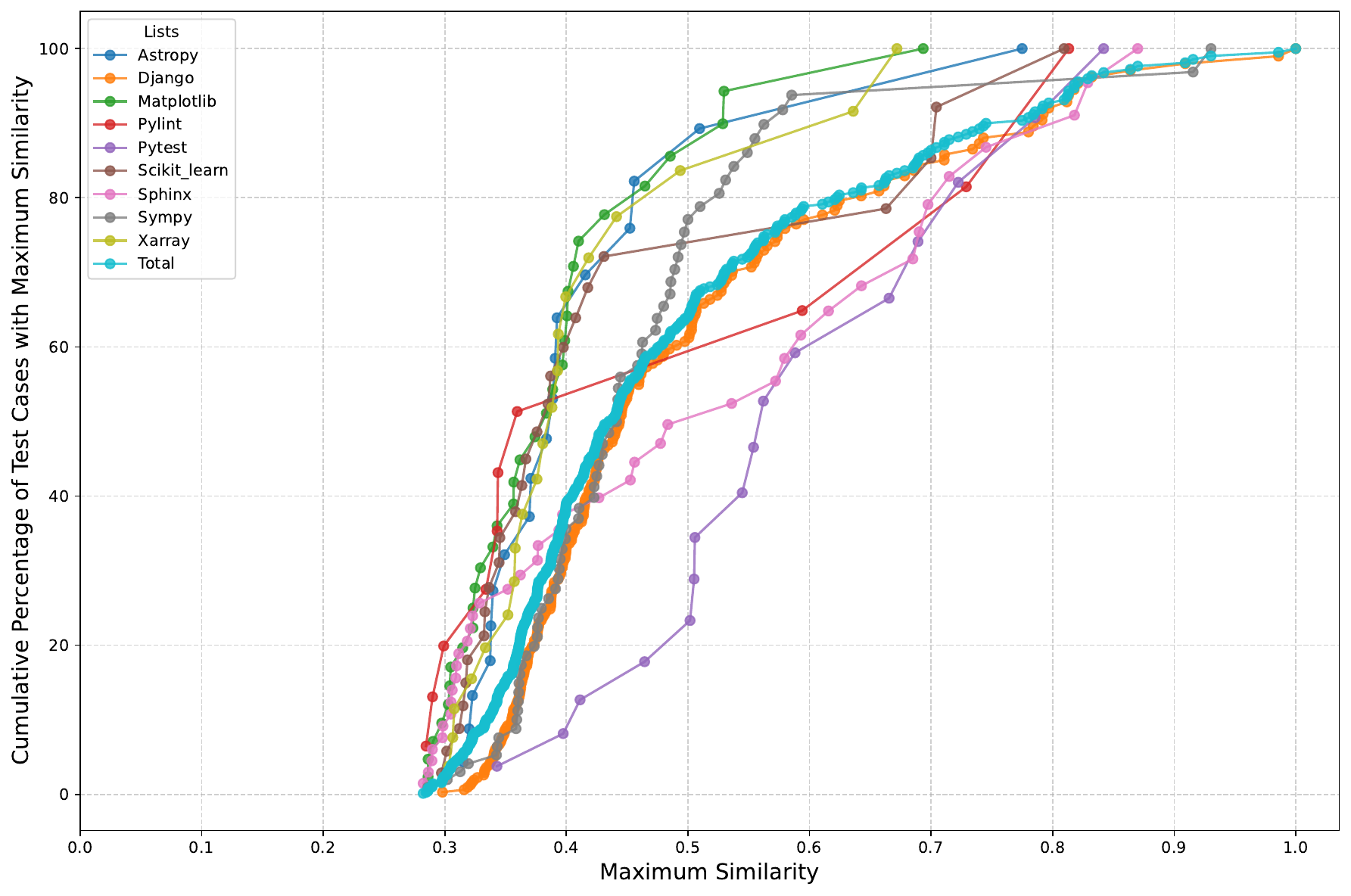}
    \caption{Cumulative percent of \solx generated all tests, using GPT-4o, with maximum similarity less than the similarity value shown on the x-axis.}
    \label{fig:overall_contamination}
    \end{minipage}
    
        \vspace{.5 in}   
    \begin{minipage}[b]{0.8\textwidth}
    \centering
    \includegraphics[width=.7\columnwidth]{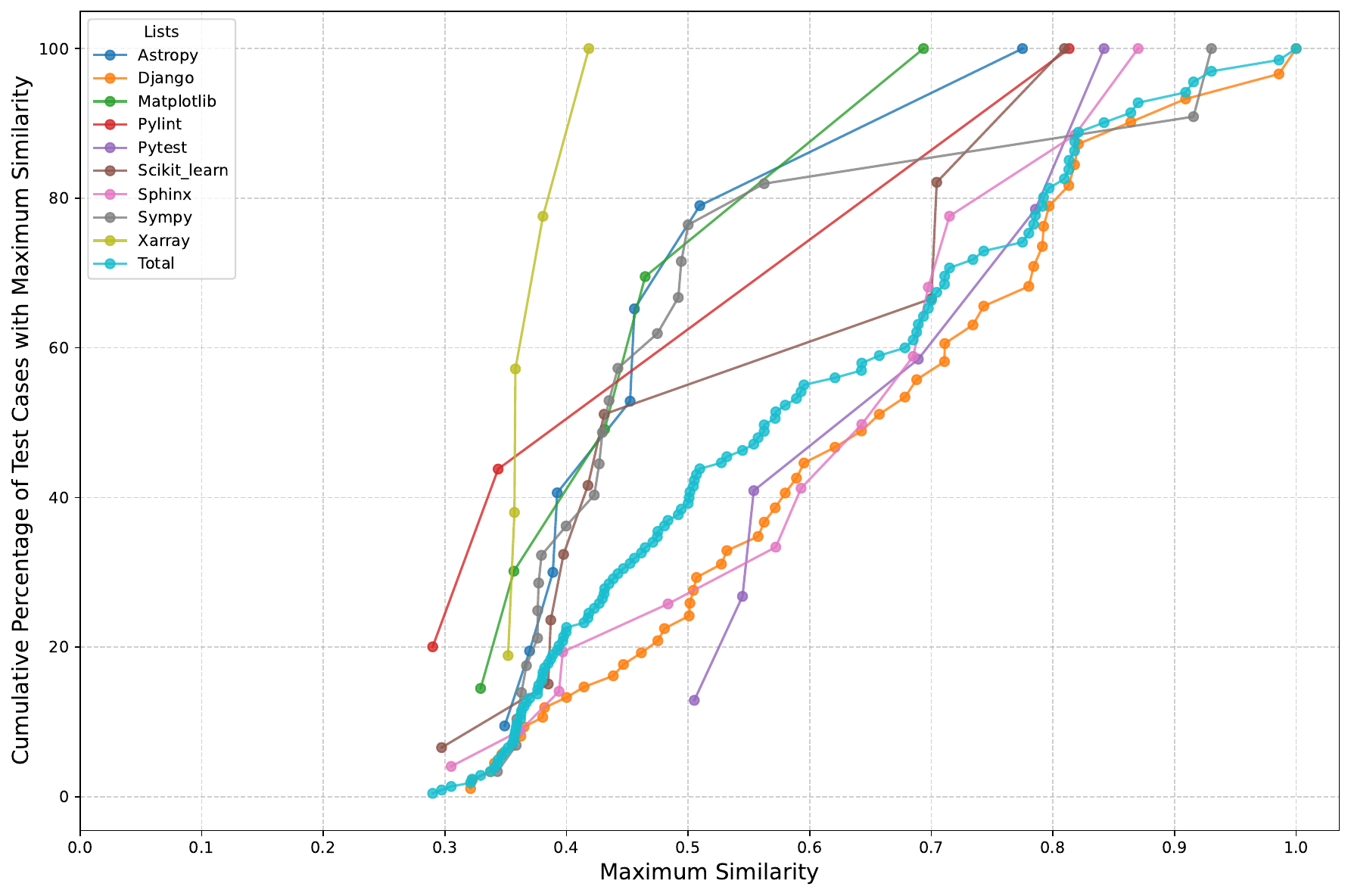}
    \caption{Cumulative percent of \solx generated modified tests, using GPT-4o, with maximum similarity less than the similarity value shown on the x-axis.}
    \label{fig:modified_contamination}
    
    \end{minipage}

\end{figure}

\begin{figure}[t!]

     \centering
    \begin{minipage}[b]{0.8\textwidth}
    \centering
    \includegraphics[width=.7\columnwidth]{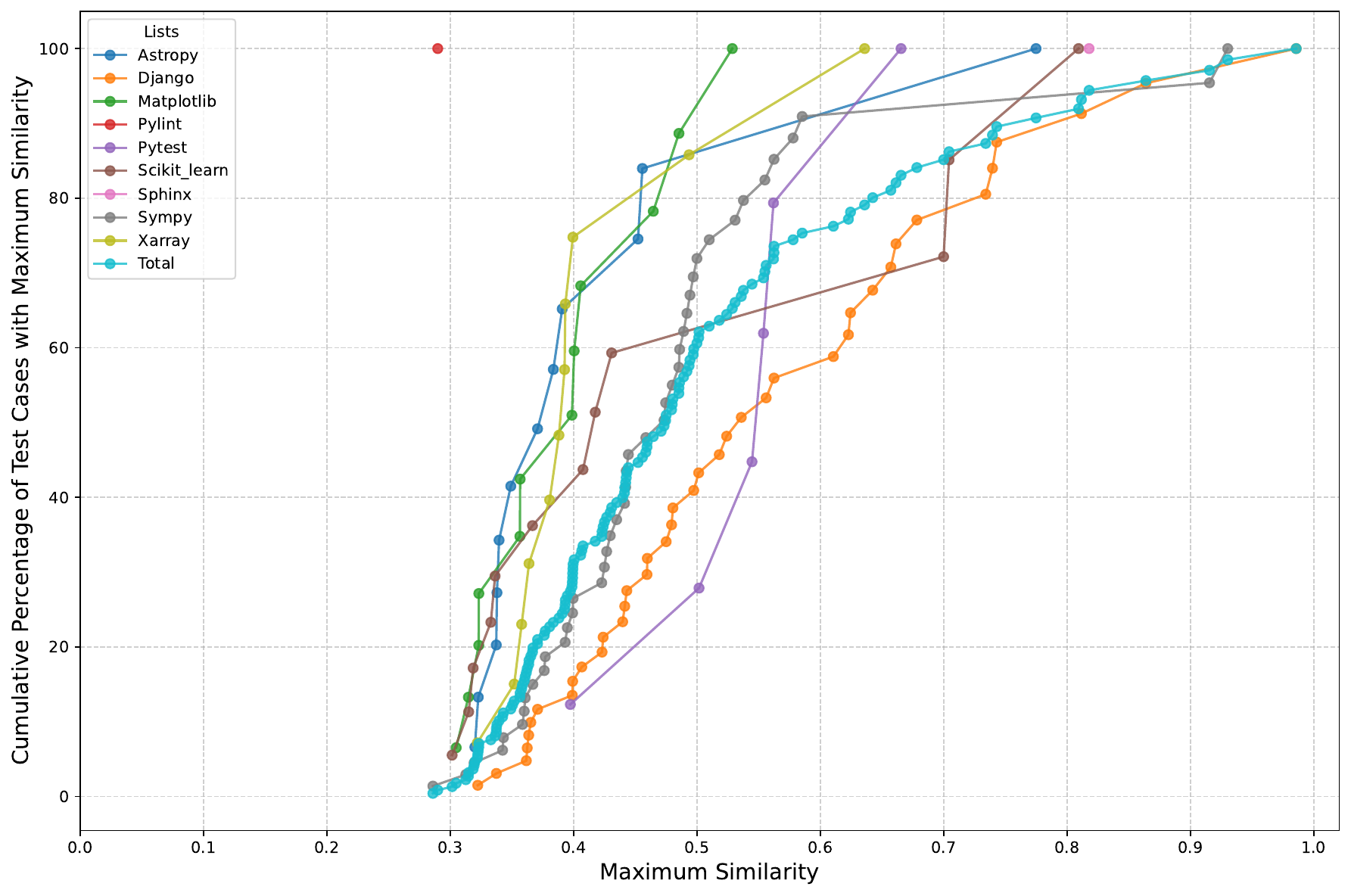}
    \caption{Cumulative percent of \solx generated fail-to-pass tests, using GPT-4o, with maximum similarity less than the similarity value shown on the x-axis.}
    \label{fig:resolved_contamination}
    
        \vspace{.5 in}   
    \end{minipage}
    \begin{minipage}[b]{0.8\textwidth}
    \centering
    \includegraphics[width=.7\columnwidth]{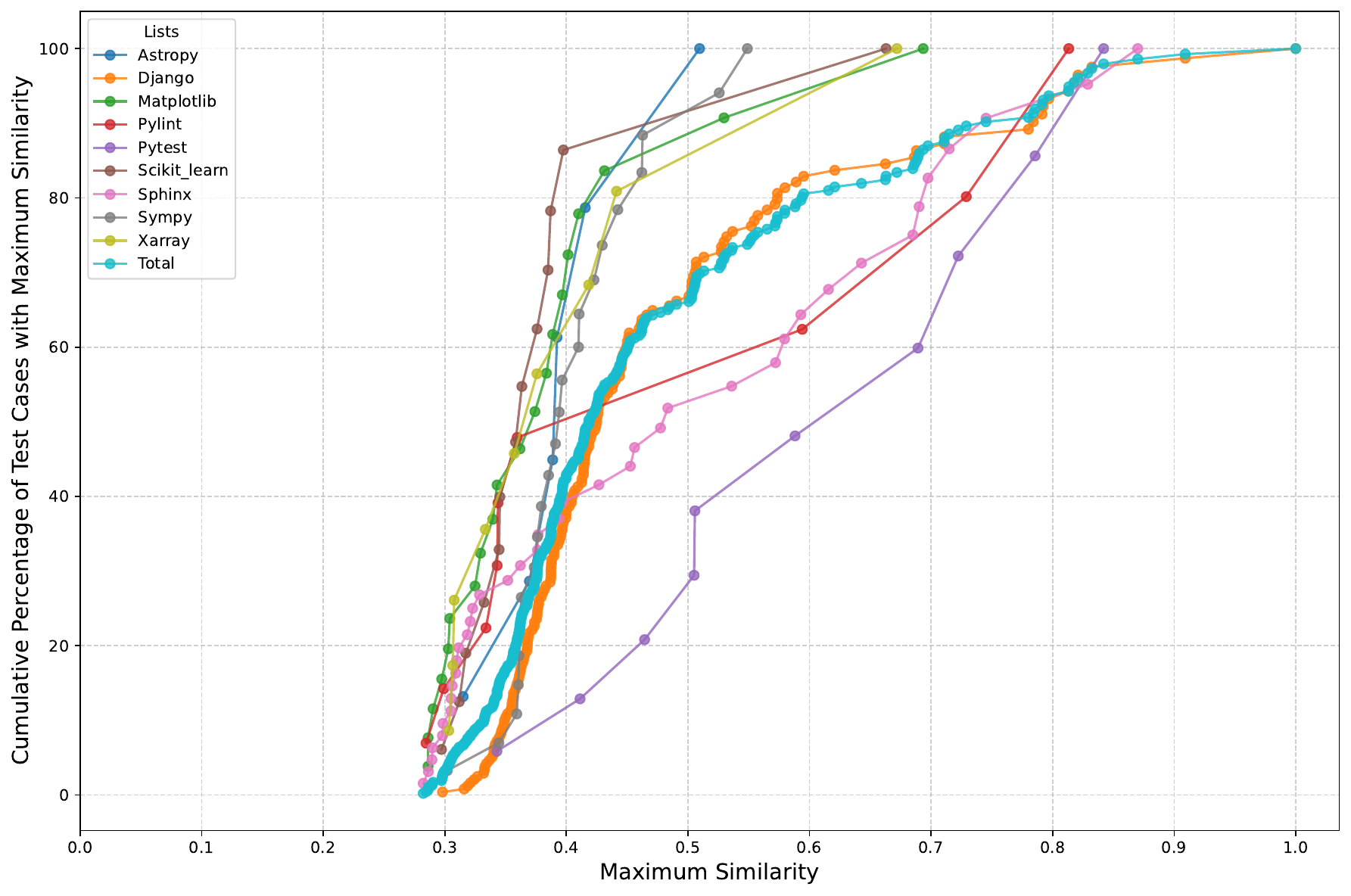}
    \caption{Cumulative percent of \solx generated non fail-to-pass tests, using GPT-4o, with maximum similarity less than the similarity value shown on the x-axis.}
    \label{fig:unresolved_contamination}
    \end{minipage}

\end{figure}

\cref{fig:overall_contamination}  to \cref{fig:unresolved_contamination} present cumulative percent of Otter generated test with different criteria, using GPT-4o, with maximum similarity less than the similarity value shown on the x-axis. \cref{tbl:similarity_example} shows examples with higher ($>0.5$) similarity score.

\begin{table*}[h]
\centering
\caption{Similarity between Generated Tests \& Existing Tests with Similarity Score.}
\resizebox{\textwidth}{!}{%
\renewcommand{\arraystretch}{1.2}% Tighter

\begin{tabular}{llr}

\toprule

\multicolumn{1}{c}{{\fontsize{14pt}{16pt}\selectfont Generated Test}}                                                                                                                                                                                                                                                                                                                                                                                                                                                                                                       & \multicolumn{1}{c}{{\fontsize{14pt}{16pt}\selectfont Test from Repo}}                                                                                                                                                                                                                                                                                                                                                                                                                                                                                                                                                                                                                                          & \multicolumn{1}{c}{{\fontsize{14pt}{16pt}\selectfont Similarity Score}} \\ \midrule

\begin{tabular}[c]{@{}l@{}}
{\fontsize{14pt}{16pt}\selectfont\texttt{def test\_call\_command\_with\_mutually\_exclusive\_group(self):}} \\     
\hspace{1em}{\fontsize{14pt}{16pt}\selectfont\texttt{out = StringIO()}} \\           
\hspace{1em}{\fontsize{14pt}{16pt}\selectfont\texttt{with self.assertRaises(CommandError) as cm:}} \\   
\hspace{2em}{\fontsize{14pt}{16pt}\selectfont\texttt{management.call\_command('my\_command', shop\_id=1, stdout=out)}} \\   
\hspace{2em}{\fontsize{14pt}{16pt}\selectfont\texttt{self.assertIn("one of the arguments --shop-id --shop is required", str(cm.exception))}} \\   
\end{tabular}                                                                                                                                                                  &

\begin{tabular}[c]{@{}l@{}}

{\fontsize{14pt}{16pt}\selectfont\texttt{def test\_language\_preserved(self):}} \\         
\hspace{1em}{\fontsize{14pt}{16pt}\selectfont\texttt{out = StringIO() }} \\        
\hspace{1em}{\fontsize{14pt}{16pt}\selectfont\texttt{with translation.override('fr'):}} \\         
\hspace{2em}{\fontsize{14pt}{16pt}\selectfont\texttt{management.call\_command('dance', stdout=out)}} \\       
\hspace{2em}{\fontsize{14pt}{16pt}\selectfont\texttt{self.assertEqual(translation.get\_language(), 'fr')}} \\

\end{tabular}

& {\fontsize{14pt}{16pt}\selectfont0.50}                                  \\ \midrule

\begin{tabular}[c]{@{}l@{}}
{\fontsize{14pt}{16pt}\selectfont\texttt{def test\_ordering\_with\_related\_field\_pk(self):}} \\ 
\hspace{1em}        {\fontsize{14pt}{16pt}\selectfont\texttt{class RelatedModel(models.Model):}} \\ 
\hspace{2em}         {\fontsize{14pt}{16pt}\selectfont\texttt{pass}} \\            
\hspace{1em}        {\fontsize{14pt}{16pt}\selectfont\texttt{class TestModel(models.Model):}} \\ 
\hspace{2em}            {\fontsize{14pt}{16pt}\selectfont\texttt{related = models.ForeignKey(RelatedModel, on\_delete=models.CASCADE)}} \\ 
\hspace{2em}             {\fontsize{14pt}{16pt}\selectfont\texttt{class Meta:}} \\ 
\hspace{3em}             	{\fontsize{14pt}{16pt}\selectfont\texttt{ordering = {[}'related\_\_pk'{]}}} \\ 
 \hspace{1em}        {\fontsize{14pt}{16pt}\selectfont\texttt{try:}} \\ 
\hspace{2em}                  {\fontsize{14pt}{16pt}\selectfont\texttt{TestModel.check()}} \\ 
\hspace{1em}         {\fontsize{14pt}{16pt}\selectfont\texttt{except ValidationError as e:}} \\ 
\hspace{2em}                 {\fontsize{14pt}{16pt}\selectfont\texttt{self.fail(f"ValidationError raised: \{e\}")}} \\ 
\end{tabular}

& 

\begin{tabular}[c]{@{}l@{}}

    {\fontsize{14pt}{16pt}\selectfont\texttt{def test\_ordering\_pointing\_to\_foreignkey\_field(self):}} \\ 
\hspace{1em}         {\fontsize{14pt}{16pt}\selectfont\texttt{class Parent(models.Model):}} \\ 
\hspace{2em}                 {\fontsize{14pt}{16pt}\selectfont\texttt{pass}} \\ 
                    
\hspace{1em}         {\fontsize{14pt}{16pt}\selectfont\texttt{class Child(models.Model):}} \\ 
\hspace{2em}                {\fontsize{14pt}{16pt}\selectfont\texttt{parent = models.ForeignKey(Parent, models.CASCADE)}} \\ 
                                       
\hspace{2em}           	{\fontsize{14pt}{16pt}\selectfont\texttt{class Meta:}} \\ 
\hspace{3em}                         {\fontsize{14pt}{16pt}\selectfont\texttt{ordering = ('parent\_id',)}} \\ 
                                                                   
\hspace{1em} {\fontsize{14pt}{16pt}\selectfont\texttt{self.assertFalse(Child.check())}} \\

\end{tabular}       

                                                                                                                                                                                                                                                                         & {\fontsize{14pt}{16pt}\selectfont0.55}                                 \\ \midrule

\begin{tabular}[c]{@{}l@{}}

    {\fontsize{14pt}{16pt}\selectfont\texttt{def test\_bulk\_update\_return\_value(self):}} \\ 
\hspace{1em}          {\fontsize{14pt}{16pt}\selectfont\texttt{for note in self.notes:}} \\ 
\hspace{2em}                      {\fontsize{14pt}{16pt}\selectfont\texttt{note.note = 'test-\%s' \% note.id}} \\ 
\hspace{2em}                      {\fontsize{14pt}{16pt}\selectfont\texttt{with self.assertNumQueries(1):}} \\ 
\hspace{3em}                              {\fontsize{14pt}{16pt}\selectfont\texttt{updated\_count = Note.objects.bulk\_update(self.notes, {[}'note'{]})}} \\ 
\hspace{3em}                               {\fontsize{14pt}{16pt}\selectfont\texttt{self.assertEqual(updated\_count, len(self.notes))}} \\

\end{tabular}                                                                                                                                                                           &

 \begin{tabular}[c]{@{}l@{}}
 
    {\fontsize{14pt}{16pt}\selectfont\texttt{def test\_simple(self):}} \\ 
\hspace{1em}         {\fontsize{14pt}{16pt}\selectfont\texttt{for note in self.notes:}} \\ 
\hspace{2em}                 {\fontsize{14pt}{16pt}\selectfont\texttt{note.note = 'test-\%s' \% note.id}} \\ 
\hspace{2em}                 {\fontsize{14pt}{16pt}\selectfont\texttt{with self.assertNumQueries(1):}} \\ 
\hspace{3em}                  		{\fontsize{14pt}{16pt}\selectfont\texttt{Note.objects.bulk\_update(self.notes, {[}'note'{]})}} \\ 
\hspace{3em}                     	{\fontsize{14pt}{16pt}\selectfont\texttt{self.assertCountEqual(}} \\ 
\hspace{4em}                                    {\fontsize{14pt}{16pt}\selectfont\texttt{Note.objects.values\_list('note', flat=True),}} \\ 
\hspace{4em}                                    {\fontsize{14pt}{16pt}\selectfont\texttt{{[}cat.note for cat in self.notes{]}}} \\ 
\hspace{3em}                            {\fontsize{14pt}{16pt}\selectfont\texttt{)}} \\

                                                                                   \end{tabular}

&

{\fontsize{14pt}{16pt}\selectfont0.61}                                 \\ \midrule

\begin{tabular}[c]{@{}l@{}}

    {\fontsize{14pt}{16pt}\selectfont\texttt{def test\_composed\_queries\_with\_values\_list(self):}} \\
\hspace{1em}         {\fontsize{14pt}{16pt}\selectfont\texttt{ReservedName.objects.create(name='a', order=2)}} \\
\hspace{1em}         {\fontsize{14pt}{16pt}\selectfont\texttt{qs1 = ReservedName.objects.all()}} \\
\hspace{1em}         {\fontsize{14pt}{16pt}\selectfont\texttt{result1 = qs1.union(qs1).values\_list('name', 'order').get()}} \\
\hspace{1em}         {\fontsize{14pt}{16pt}\selectfont\texttt{self.assertEqual(result1, ('a', 2))}} \\
\hspace{1em}         {\fontsize{14pt}{16pt}\selectfont\texttt{result2 = qs1.union(qs1).values\_list('order').get()}} \\
\hspace{1em}         {\fontsize{14pt}{16pt}\selectfont\texttt{self.assertEqual(result2, (2,))}} \\

\end{tabular}

  & \begin{tabular}[c]{@{}l@{}}
  
    {\fontsize{14pt}{16pt}\selectfont\texttt{def test\_union\_with\_values(self):}} \\
\hspace{1em}         {\fontsize{14pt}{16pt}\selectfont\texttt{ReservedName.objects.create(name='a', order=2)}} \\
\hspace{1em}        {\fontsize{14pt}{16pt}\selectfont\texttt{qs1 = ReservedName.objects.all()}} \\
\hspace{1em}         {\fontsize{14pt}{16pt}\selectfont\texttt{reserved\_name = qs1.union(qs1).values('name', 'order', 'id').get()}} \\
\hspace{1em}         {\fontsize{14pt}{16pt}\selectfont\texttt{self.assertEqual(reserved\_name{[}'name'{]}, 'a')}} \\
\hspace{1em}        {\fontsize{14pt}{16pt}\selectfont\texttt{self.assertEqual(reserved\_name{[}'order'{]}, 2)}} \\
\hspace{1em}        {\fontsize{14pt}{16pt}\selectfont\texttt{reserved\_name = qs1.union(qs1).values\_list('name', 'order', 'id').get()}} \\
\hspace{1em}         {\fontsize{14pt}{16pt}\selectfont\texttt{self.assertEqual(reserved\_name{[}:2{]}, ('a', 2))}} \\

  \end{tabular}

          & {\fontsize{14pt}{16pt}\selectfont0.66}                                 \\ \midrule

\begin{tabular}[c]{@{}l@{}}

    {\fontsize{14pt}{16pt}\selectfont\texttt{def test\_type\_hints\_in\_uml\_generation(project):}} \\
\hspace{1em}	{\fontsize{14pt}{16pt}\selectfont\texttt{klass = project.get\_module("data.clientmodule\_test"){[}"C"{]}}} \\
\hspace{1em}         {\fontsize{14pt}{16pt}\selectfont\texttt{assert hasattr(klass, "instance\_attrs\_type")}} \\
\hspace{1em}         {\fontsize{14pt}{16pt}\selectfont\texttt{type\_dict = klass.instance\_attrs\_type}} \\
\hspace{1em}	{\fontsize{14pt}{16pt}\selectfont\texttt{assert len(type\_dict) == 1}} \\
\hspace{1em}	{\fontsize{14pt}{16pt}\selectfont\texttt{keys = sorted(type\_dict.keys())}} \\
\hspace{1em}	{\fontsize{14pt}{16pt}\selectfont\texttt{assert keys == {[}"a"{]}}} \\
\hspace{1em}	{\fontsize{14pt}{16pt}\selectfont\texttt{assert isinstance(type\_dict{[}"a"{]}{[}0{]}, astroid.bases.Instance), type\_dict{[}"a"{]}}} \\
\hspace{1em}	{\fontsize{14pt}{16pt}\selectfont\texttt{assert type\_dict{[}"a"{]}{[}0{]}.name == "str"}} \\

\end{tabular}

 &

 \begin{tabular}[c]{@{}l@{}}

    {\fontsize{14pt}{16pt}\selectfont\texttt{def test\_instance\_attrs\_resolution(project):}} \\
\hspace{1em}	{\fontsize{14pt}{16pt}\selectfont\texttt{klass = project.get\_module("data.clientmodule\_test"){[}"Specialization"{]}}} \\
\hspace{1em}	{\fontsize{14pt}{16pt}\selectfont\texttt{assert hasattr(klass, "instance\_attrs\_type")}} \\
\hspace{1em}	{\fontsize{14pt}{16pt}\selectfont\texttt{type\_dict = klass.instance\_attrs\_type}} \\
\hspace{1em}	{\fontsize{14pt}{16pt}\selectfont\texttt{assert len(type\_dict) == 2}} \\
\hspace{1em}	{\fontsize{14pt}{16pt}\selectfont\texttt{keys = sorted(type\_dict.keys())}} \\
\hspace{1em}	{\fontsize{14pt}{16pt}\selectfont\texttt{assert keys == {[}"\_id", "relation"{]}}} \\
\hspace{1em}	{\fontsize{14pt}{16pt}\selectfont\texttt{assert isinstance(type\_dict{[}"relation"{]}{[}0{]}, astroid.bases.Instance), type\_dict{[}}} \\
\hspace{2em}		{\fontsize{14pt}{16pt}\selectfont\texttt{"relation"}} \\
\hspace{1em}        {\fontsize{14pt}{16pt}\selectfont\texttt{{]}}} \\
\hspace{1em}	{\fontsize{14pt}{16pt}\selectfont\texttt{assert type\_dict{[}"relation"{]}{[}0{]}.name == "DoNothing"}} \\
\hspace{1em}         {\fontsize{14pt}{16pt}\selectfont\texttt{assert type\_dict{[}"\_id"{]}{[}0{]} is astroid.Uninferable}} \\

 \end{tabular}

  & {\fontsize{14pt}{16pt}\selectfont0.73}                                 \\ \midrule

\begin{tabular}[c]{@{}l@{}}

    {\fontsize{14pt}{16pt}\selectfont\texttt{def test\_classmethod\_property(app):}} \\
\hspace{1em}      {\fontsize{14pt}{16pt}\selectfont\texttt{actual = do\_autodoc(app, 'property', 'target.properties.Foo.classmethod\_prop')}} \\
\hspace{1em}      {\fontsize{14pt}{16pt}\selectfont\texttt{assert list(actual) == {[}}} \\
\hspace{2em}                    {\fontsize{14pt}{16pt}\selectfont\texttt{'',}} \\
\hspace{2em}                           {\fontsize{14pt}{16pt}\selectfont\texttt{'.. py:property:: Foo.classmethod\_prop',}} \\
\hspace{2em}                                   {\fontsize{14pt}{16pt}\selectfont\texttt{'   :module: target.properties',}} \\
\hspace{2em}                                             {\fontsize{14pt}{16pt}\selectfont\texttt{'   :type: str',}} \\
\hspace{2em}                                                     {\fontsize{14pt}{16pt}\selectfont\texttt{'',}} \\
\hspace{2em}                                                              {\fontsize{14pt}{16pt}\selectfont\texttt{'   Some class property.',}} \\
\hspace{2em}                                                                       {\fontsize{14pt}{16pt}\selectfont\texttt{'',}} \\
\hspace{1em}                                                                           {\fontsize{14pt}{16pt}\selectfont\texttt{{]}}} \\

\end{tabular}                                                                                                                 &

\begin{tabular}[c]{@{}l@{}}

    {\fontsize{14pt}{16pt}\selectfont\texttt{def test\_properties(app):}} \\
\hspace{1em}     {\fontsize{14pt}{16pt}\selectfont\texttt{actual = do\_autodoc(app, 'property', 'target.properties.Foo.prop')}} \\
\hspace{1em}     {\fontsize{14pt}{16pt}\selectfont\texttt{assert list(actual) == {[}}} \\
\hspace{2em}                   {\fontsize{14pt}{16pt}\selectfont\texttt{'',}} \\
\hspace{2em}                            {\fontsize{14pt}{16pt}\selectfont\texttt{'.. py:property:: Foo.prop',}} \\
\hspace{2em}                                    {\fontsize{14pt}{16pt}\selectfont\texttt{'   :module: target.properties',}} \\
\hspace{2em}                                            {\fontsize{14pt}{16pt}\selectfont\texttt{'   :type: int',}} \\
\hspace{2em}                                                    {\fontsize{14pt}{16pt}\selectfont\texttt{'',}} \\
\hspace{2em}                                                            {\fontsize{14pt}{16pt}\selectfont\texttt{'   docstring',}} \\
\hspace{2em}                                                                    {\fontsize{14pt}{16pt}\selectfont\texttt{'',}} \\
\hspace{1em}                                                                        {\fontsize{14pt}{16pt}\selectfont\texttt{{]}}} \\

\end{tabular}

                                                                                                                                                                                                                                                                                       & {\fontsize{14pt}{16pt}\selectfont0.82}            \\ \bottomrule                     
\end{tabular}

}
\label{tbl:similarity_example}
\end{table*}

%\begin{figure}[h]
%    \centering
%    \includegraphics[width=.6\columnwidth]{overall_contamination.pdf}
%    \caption{Cumulative percent of \solx generated all tests, using GPT-4o, with maximum similarity less than the similarity value shown on the x-axis.}
%    \label{fig:overall_contamination}
%\end{figure}
%
%
%\begin{figure}[h]
%    \centering
%    \includegraphics[width=.6\columnwidth]{modified_contamination.pdf}
%    \caption{Cumulative percent of \solx generated modified tests, using GPT-4o, with maximum similarity less than the similarity value shown on the x-axis.}
%    \label{fig:modified_contamination}
%\end{figure}
%
%
%\begin{figure}[h]
%    \centering
%    \includegraphics[width=.6\columnwidth]{resolved_contamination.pdf}
%    \caption{Cumulative percent of \solx generated fail-to-pass tests, using GPT-4o, with maximum similarity less than the similarity value shown on the x-axis.}
%    \label{fig:resolved_contamination}
%\end{figure}
%
%\begin{figure}[h]
%    \centering
%    \includegraphics[width=.6\columnwidth]{unresolved_contamination.pdf}
%    \caption{Cumulative percent of \solx generated non fail-to-pass tests, using GPT-4o, with maximum similarity less than the similarity value shown on the x-axis.}
%    \label{fig:unresolved_contamination}
%\end{figure}

\section{Prompts for \solx}

\begin{figure}[t!]
    \centering
    \begin{minipage}[b]{0.8\textwidth}
        \centering
        \includegraphics[width=\textwidth]{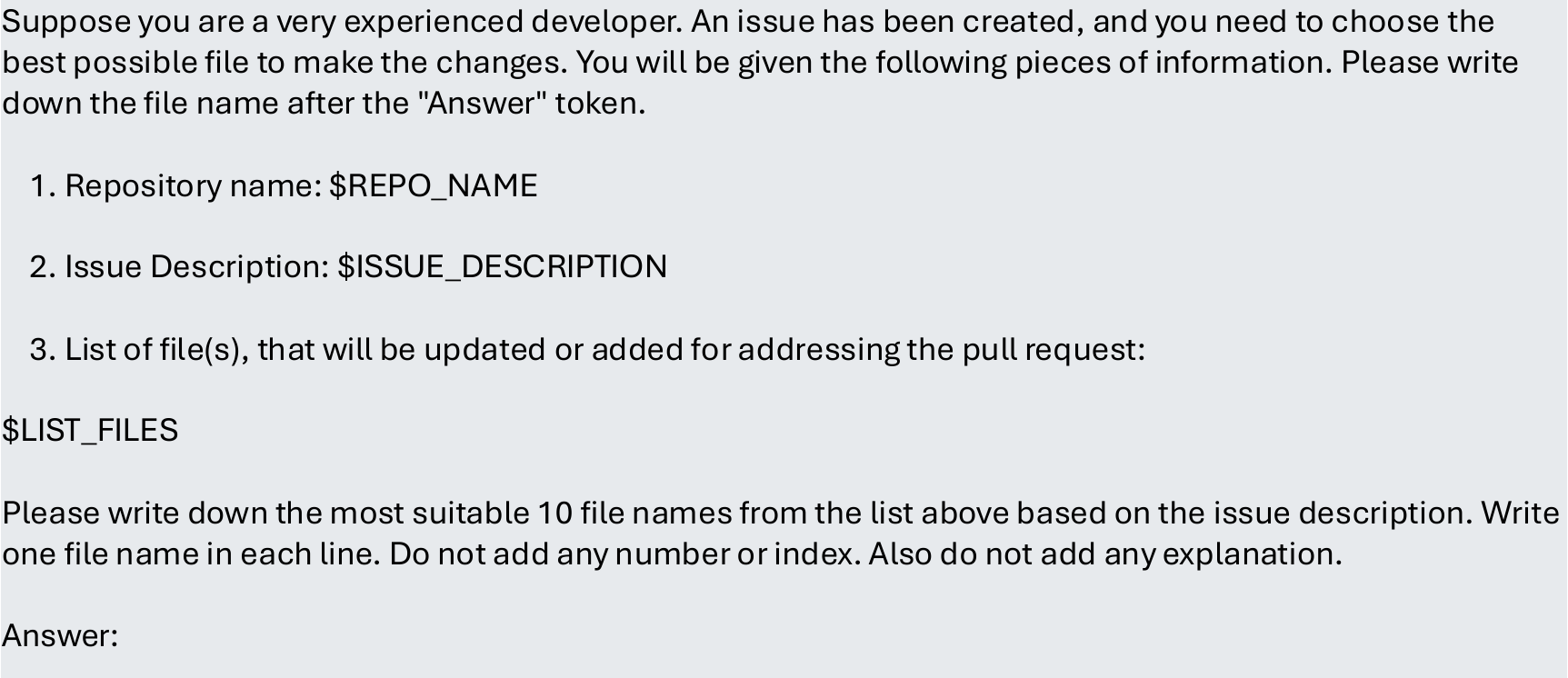}

        \caption{Prompt (1 of 2) for focal function localizer.}
                \label{fig:focal1}
    \end{minipage}

    \vspace{.5 in}
    
    \begin{minipage}[b]{0.8\textwidth}
        \centering
        \includegraphics[width=\textwidth]{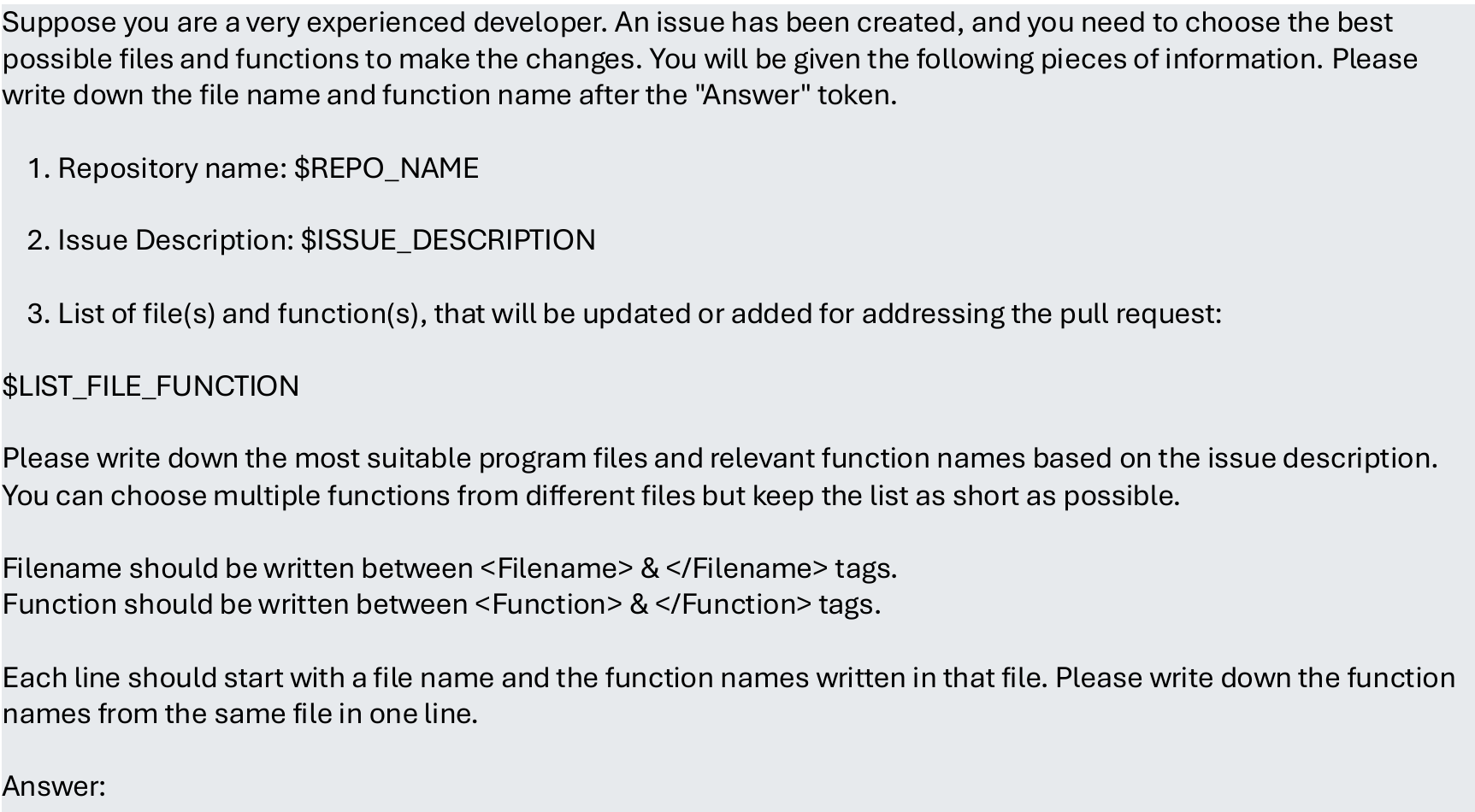}
        \caption{Prompt (2 of 2) for focal function localizer.}
         \label{fig:focal11}
    \end{minipage}
    
%\caption{Prompts for focal function localizer. Using first prompt, we localize a list of files and second prompt, we localize the focal functions.}
%\label{fig:focal_prompt}    

\end{figure}

\begin{figure}[t!]
    \centering
    \begin{minipage}[b]{0.8\textwidth}
        \centering
        \includegraphics[width=\textwidth]{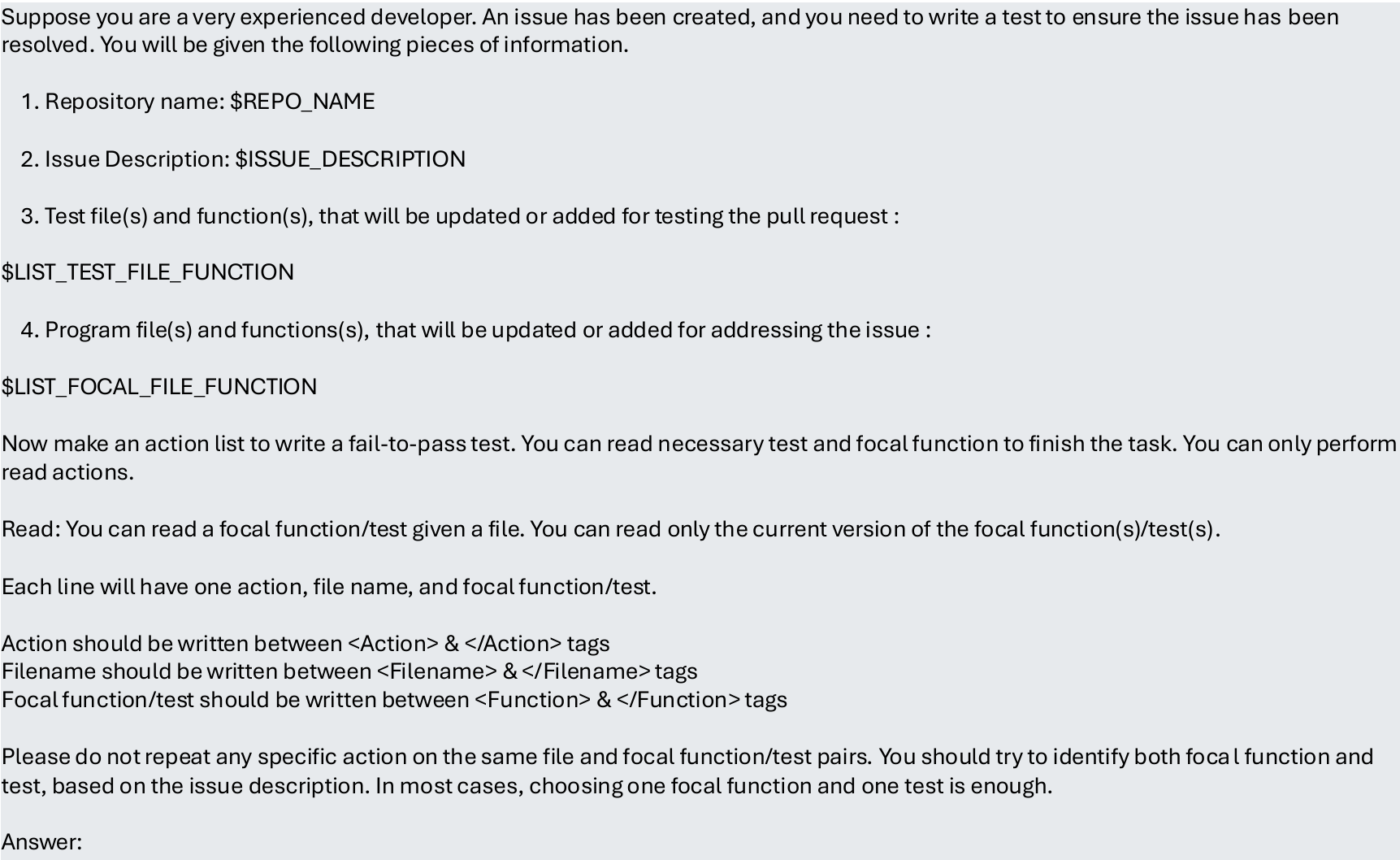}
        \caption{Prompts for ``make an initial plan'' step.}
        \label{plan1}
    \end{minipage}
    
            \vspace{.5 in}   
    
    \begin{minipage}[b]{1\textwidth}
        \centering
        \includegraphics[width=\textwidth]{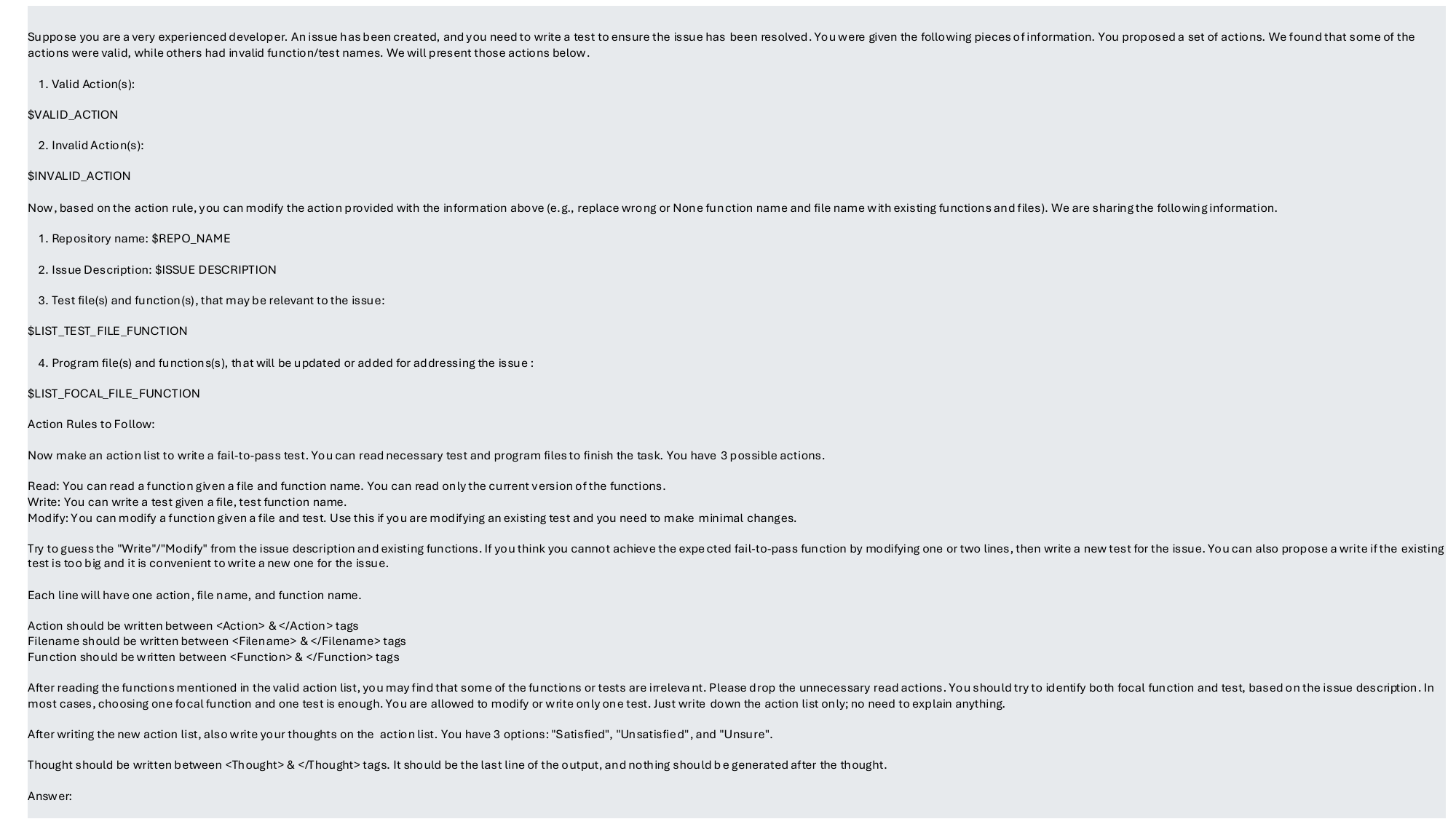}
        \caption{Prompts for ``reflect and improve the plan'' step.}
        \label{plan2}
    \end{minipage}
    
%\caption{Prompts for self-reflective action planning.}
%\label{fig:action_plan}      

\end{figure}

\begin{figure}[t!]
    \centering
    \begin{minipage}[b]{0.8\textwidth}
        \centering
        \includegraphics[width=\textwidth]{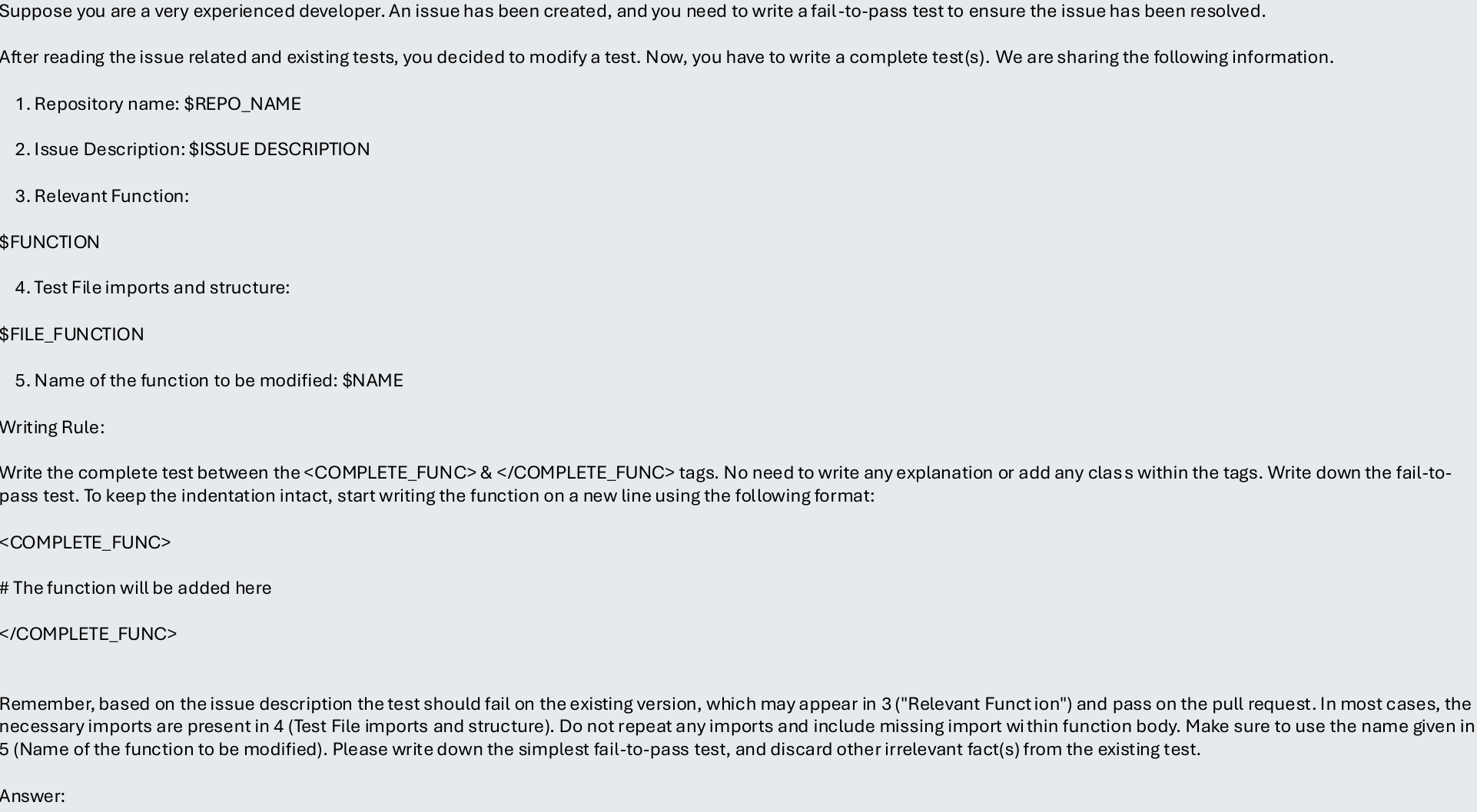}
        \caption{Prompt for modifying existing test.}
        \label{modifyprompt}
    \end{minipage}
    
            \vspace{.5 in}   
    
    \begin{minipage}[b]{.8\textwidth}
        \centering
        \includegraphics[width=\textwidth]{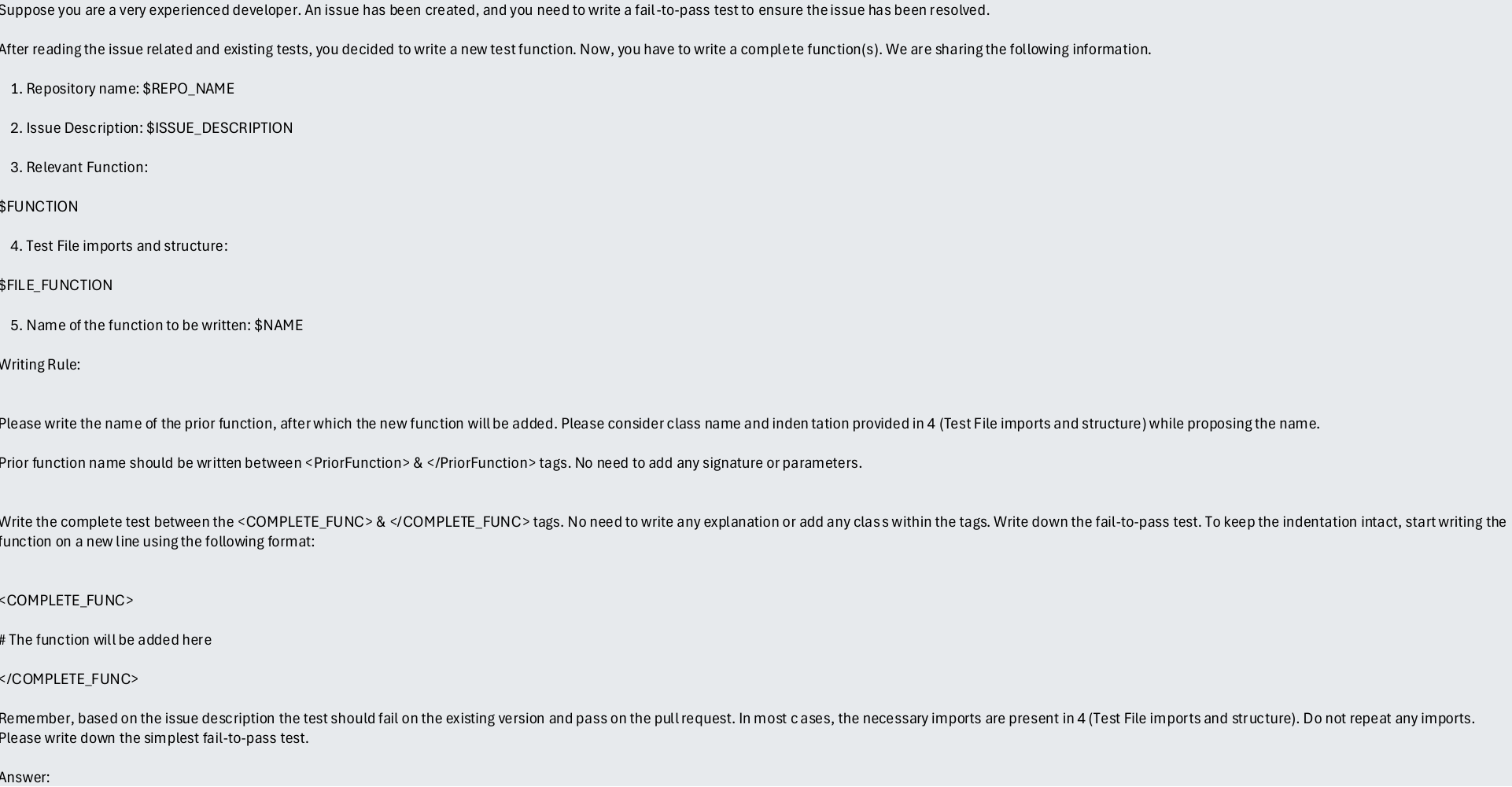}
        \caption{Prompt for writing new test.}
        \label{fig:writeprompt}
    \end{minipage}
\end{figure}

\cref{fig:focal1} to \cref{fig:writeprompt} present all the prompts used for \solx. 

\begin{figure}[t!]
    \centering
    \begin{minipage}[b]{0.8\textwidth}
        \centering
        \includegraphics[width=\textwidth]{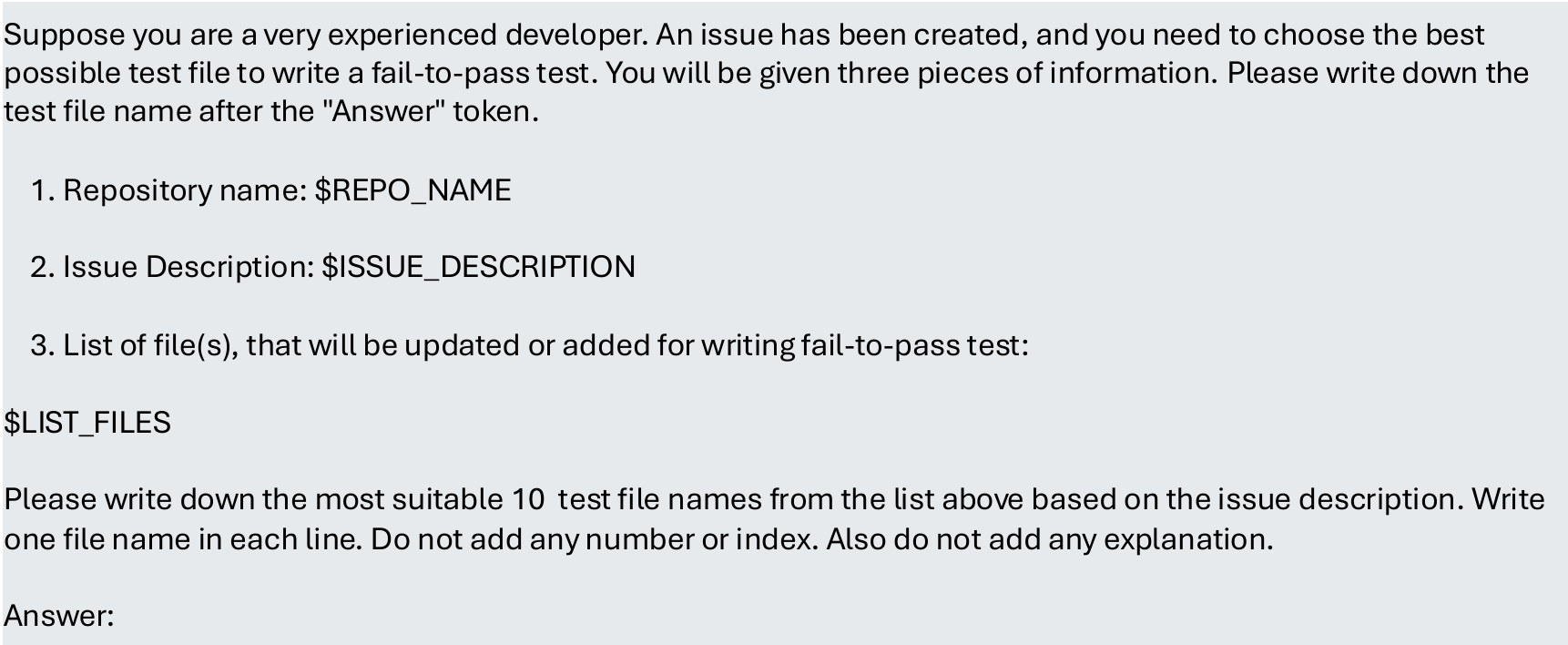}
 \caption{Prompt (1 of 2) for test function localizer.}

    \end{minipage}
    
    \vspace{.5 in}

    \begin{minipage}[b]{0.8\textwidth}
        \centering
        \includegraphics[width=\textwidth]{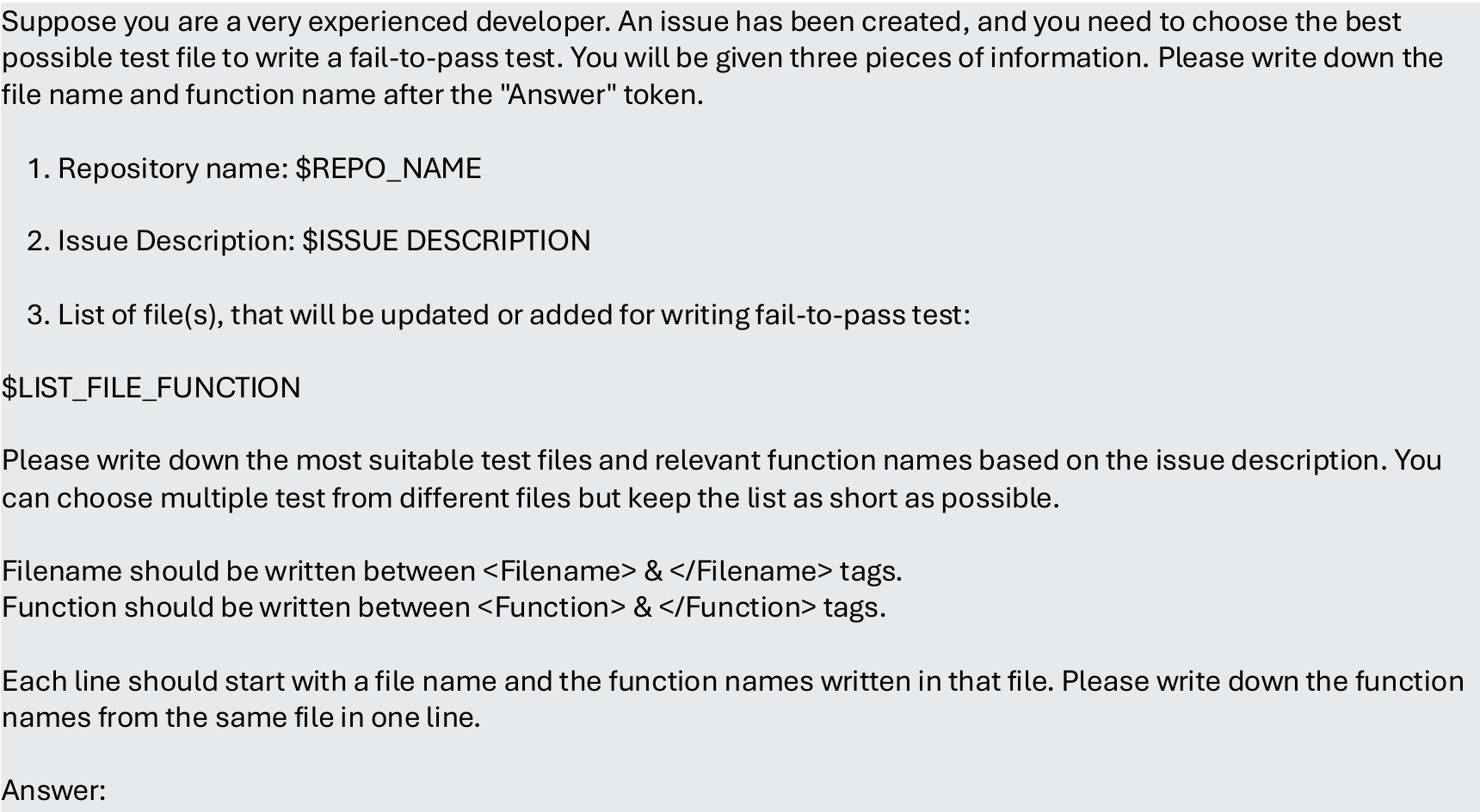}
        \caption{Prompt (2 of 2) for test function localizer.}
    \end{minipage}
    
%\caption{Prompts for test function localizer. Using first prompt, we localize a list of files and second prompt, we localize the test functions.}
%\label{fig:focal_prompt}    

\end{figure}

%The $\mathtt{\backslash onecolumn}$ command above can be kept in place if you prefer a one-column appendix, or can be removed if you prefer a two-column appendix.  Apart from this possible change, the style (font size, spacing, margins, page numbering, etc.) should be kept the same as the main body.
%%%%%%%%%%%%%%%%%%%%%%%%%%%%%%%%%%%%%%%%%%%%%%%%%%%%%%%%%%%%%%%%%%%%%%%%%%%%%%%
%%%%%%%%%%%%%%%%%%%%%%%%%%%%%%%%%%%%%%%%%%%%%%%%%%%%%%%%%%%%%%%%%%%%%%%%%%%%%%%

\end{document}

%% file: intro.tex
\section{Introduction}\label{sec:intro}

A software engineering~(SWE) \emph{issue} is a bug report or feature
request for improving the code in a repository.
Before a software engineer attempts to write new code that resolves a
bug, they typically write a \emph{reproduction test} to confirm the
presence of the bug in the old code.
In fact, even when working on a feature request, before writing new
code it is recommended to create an \emph{acceptance test} first, to
confirm the presence of the feature in the new code once written.
This practice is known as test-driven development~(TDD), and it
improves the quality of both tests and the
code itself~\cite{beck_2002}.
The recent introduction of SWE-Bench~\cite{jimenezswe} has spurred
work on resolving SWE issues automatically with LLMs, typically with
agents~\cite{sweagent2,autocoderover2,opendevin}.
Such issue-resolution systems are often called \emph{SWE agents}.
We posit that, just like up-front tests
help human SWE engineers, they also help automated SWE agents.

% Very recently, there are some initial solutions, such as
% \mbox{SWE-Agent+~\cite{mundler2024swtbench}}.
% Unfortunately, they are expensive (require many LLM calls and tokens)
% and have limited success (e.g.\ the state-of-the-art SWE-Agent+ only
% generates a fail-to-pass test in 19.2\% of cases).

While test generation is an active area of research, prior work focuses
on creating tests for existing code, not on creating tests from
issue descriptions alone. A recent solution \mbox{SWE-Agent+~\cite{mundler2024swtbench}} attempts to create tests from issue descriptions albeit with limited success (the state-of-the-art
SWE-Agent+ only generates a fail-to-pass test in 19.2\% cases and is quite expensive, requiring several LLM calls
and tokens). See Section~\ref{sec:related} for a more in-depth discussion of
related work.

This paper introduces \solx (acronym for ``\solx: TDD-Test gEnerator
for Reproducing issues'').
It takes as input $x$ an issue description and the original code
in the repository, and generates as output a set of tests~$y$.
\solx contains a novel self-reflective action planner for deciding
which code to read and which tests to write.
It also uses rule-based code analyses and transformations throughout
the workflow to curate LLM inputs, validate LLM outputs, and repair
generated tests.
\solx generates fail-to-pass tests in 31.4\% of cases, and in an
ensemble of size~5 dubbed \soly, that increases to 37.0\%.
At the same time, the cost (even with ensembling) is less than \$0.10
per issue with GPT-4o.
Tests generated by \solx adhere to the same testing framework as
repository's existing test suite, to which they can be added.

Today the most popular benchmark for automatically resolving SWE
issues is SWE-bench Verified~\cite{chowdhury_et_al_2024}.
To evaluate solutions such as \solx that automatically generate tests
from issues, we introduce a new benchmark, \tdd.
\tdd evaluates tests by checking whether the tests a) fail on the old
code before issue resolution, b) pass on the new code, and
c) cover the code changes well.
% It evaluates tests by checking whether they fail on the old code
% before issue resolution, pass on the new code, and cover the code
% changes well.
The fact that \tdd is derived from SWE-bench Verified enables us to
empirically study the effects of generated tests on SWE agents.
We observe that the tests from \soly can be used to trade precision for
recall on the SWE-bench Verified leaderboard.
For instance, for the system ranked 3rd on the leaderboard, filtering by the
generated tests boosts precision from 60.8\% to 91.9\% while
reducing recall to 33\%.
This paper makes the following contributions:

\begin{itemize}[leftmargin=1em, topsep=0pt, itemsep=0pt, parsep=0pt]
  \item \solx, a system that generates tests from issues, using LLMs
    with a novel self-reflective action planning technique along with
    rule-based pre- and post-processing.
  \item \tdd\footnote{\url{https://github.com/IBM/TDD-Bench-Verified}}, a benchmark for evaluating tests generated from issues,
    including a high-quality dataset and a metric based on test
    results and coverage.
  \item An empirical study on using tests generated from issues to
    filter issue-resolution candidates for SWE-bench Verified.
\end{itemize}

Generated tests can assist software engineers both before and after
resolving an issue, and can increase trust in automated
SWE agents.

%% file: problem.tex
\section{Problem Statement}\label{sec:problem}

This work focuses on the problem of generating tests from issues.
Specifically, the input $x$ is a pair
\mbox{$\langle d_\mathrm{issue},c_\mathrm{old}\rangle$}
of an issue description $d_\mathrm{issue}$ alongside the old version
$c_\mathrm{old}$ of the code before the issue is resolved.
The issue description is typically in natural language, but it may
sometimes contain embedded code snippets or stack traces.
The code is a snapshot of all the files and folders in a Python
source code repository.
The expected output $y$ is a set of tests that should go from failing
on $c_\mathrm{old}$ to passing on $c_\mathrm{new}$, which is the new
version of the code after the issue is resolved.
By failing on $c_\mathrm{old}$, the tests reproduce the issue, and by
passing on $c_\mathrm{new}$, they validate its resolution.
Besides going from failing to passing, the tests should also maximize
coverage of the code change (formalized in Section~\ref{sec:tddbench}).
A solution to this problem is thus a function $\mathit{genTests}$ that
takes an input $x$ and returns tests \mbox{$y=\mathit{genTests}(x)$}.
The new code $c_\mathrm{new}$ is not available to $\mathit{genTests}$, which must
generate tests $y$ based on $x$ alone.
This is representative of the real world where source code
repositories may have regression tests for existing code but lack
tests for open issues.
\solx provides an implementation of the $\mathit{genTests}$ function
and, thus, a solution to this problem.

%% file: relatedwork.tex
\section{Related Work}\label{sec:related}

% Ten years before the SWE-bench dataset of Python
% issues~\cite{jimenezswe} came the Defects4J dataset of Java
% issues~\cite{just2014defects4j}.

Prior to SWE-bench~(Python)~\cite{jimenezswe},
Defects4J~(Java)~\cite{just2014defects4j} has been a popular benchmark
in the community.
The creators of Defects4J carefully curated and cleaned up each issue
by hand.
Unlike SWE-bench, Defects4J only contains bug reports, no feature
requests.
The earliest system we are aware of that generates tests from issues,
Libro~\cite{kang2023large}, focuses on Defects4J.
Libro achieves a fail-to-pass rate of 19.9\% with one generation.
(\citet{mundler2024swtbench} ported Libro to Python and measured a
fail-to-pass rate of 15.2\%.)
\citet{plein_et_al_2024} proposed another test-generation system for
Defects4J, reporting a fail-to-pass rate of 6\%.
Both systems have relatively low success rates, and
unlike our work, neither evaluates the impact of generated tests on
issue-resolving systems.

When resolving issues, some SWE agents also generate tests along the
way.
% SWE-Agent is a single-agent system, prompted to start by using its
% general tools to attempt to reproduce the issue~\cite{sweagent2}.
The original
SWE-Agent~\cite{sweagent2}, a single-agent system, attempts to reproduce the issue,
as explicitly instructed in its prompt.
% And both CodeR~\cite{sweagent3coder} and
% SpecRover~\cite{autocoderover2} are multi-agent systems, starting with
% a reproducer agent for generating tests.
Some multi-agent systems---CodeR~\cite{sweagent3coder} and SpecRover~\cite{autocoderover2}---start with a Reproducer agent for generating tests.
However, none of the three (SWE-Agent, CodeR, or SpecRover)
are evaluated for the effectiveness of their generated tests.
Agentless~\cite{sweagent5agentless} relies on inference
scaling, generating several candidate patches and several
tests.
It then uses the tests to help rank the patches, ultimately choosing a
single patch to submit.
The effectiveness of the tests is evaluated indirectly by their impact
on issue resolution rate (from 27\% to 32\%), not directly for their
own fail-to-pass rate or coverage like in our work.

Three very recent systems are dedicated to generating tests from Python
issues.
Aegis~\cite{aegis} is a multi-agent system that uses inference
scaling, but the exact dataset for their evaluation
is unclear (the paper says SWE-bench Lite, but then compares against
numbers from another system on SWT-bench Lite, which is different).
Aegis is more costly than \soly, and unlike our
work, the Aegis paper (a) does not report coverage numbers and (b) does not evaluate how tests can help trade off precision vs.\ recall w.r.t. performance of SWE agents.
EvoCoder~\cite{lin2024llms} uses experiences from prior issues to help with the latest issue at hand, which is complementary to \solx. Unlike our work, the generated tests are not integrated with the existing CI pipeline. Furthermore, the experiments do not use execution-based metrics, making it hard to compare empirically.
SWE-Agent+ adapts a patch-generating agent to generate tests
instead~\cite{mundler2024swtbench} and achieves 19.2\% fail-to-pass rate on the SWT-bench Lite dataset
introduced by the same paper.
SWT-bench applies less rigorous quality filters than \tdd, and
measures coverage in a round-about way by first running additional
tests than just generated ones and then subtracting them back out.
Using tests from SWE-Agent+ as a filter improves precision
of SWE agents to 47.8\% while reducing recall to~20\%.
\soly outperforms SWE-Agent+ on all of these metrics.

CodeT~\cite{chen2023codet} is one of the first approaches that leverage the same LLM to automatically generate both code samples and corresponding test cases. CodeT then executes generated code samples using the generated test cases and performs a dual execution agreement to choose the best solution and test. CodeT is not directly applicable to our current setup because we have multiple tests but no code patch. 
CodeMonkeys~\cite{ehrlich2025codemonkeys} is another relevant work that iteratively and jointly improves a patch with a test, which is interesting and complementary to \solx. Unlike our work, their experiments do not evaluate the fail-to-pass rate nor the coverage of generated tests.

%% file: method.tex
\section{Methodology}\label{sec:method}

This section describes three solutions to the problem stated in
Section~\ref{sec:problem}: \solx, \soly, and a baseline approach.

\subsection{\solx: Test Generation Guided by Self-reflective Action Planner}

\begin{figure}[t]
  \centerline{\includegraphics[width=\columnwidth]{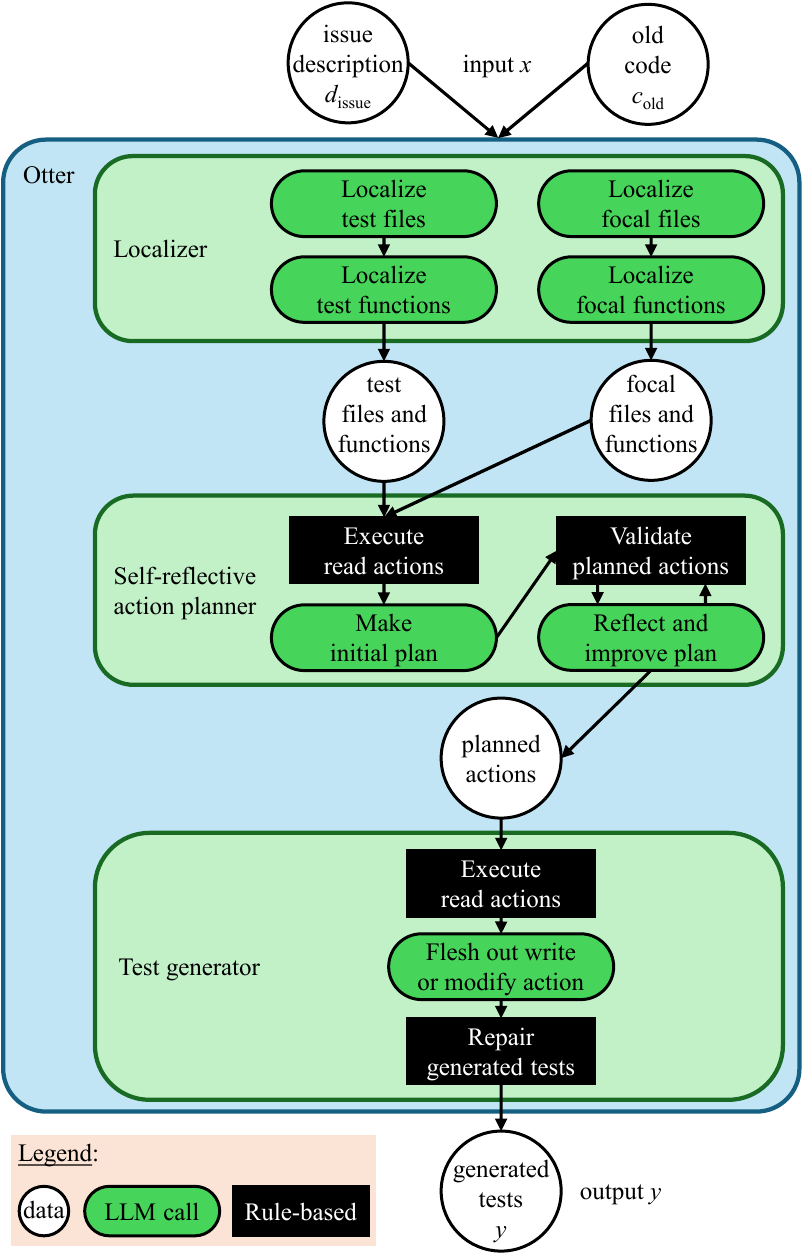}}
  \caption{\label{fig:overview}Overview of \solx.}
\end{figure}

\cref{fig:overview} presents an overview of \solx, which has three main
components: a localizer, a planner, and a test generator.

\paragraph{Localizer.}

Before developers tackle an issue, they usually start by
gaining an understanding of the current state $c_\mathrm{old}$ of the
project.
They do this by localizing relevant existing tests and \emph{focal}
functions (i.e., functions exercised by those tests and likely places
for a fix).
Depending on the developer's familiarity with the project and the
nature of the issue, the difficulty of localization may vary.
Inspired by human developers' actions, our approach also starts with
test and focal function localization.
The localizer phase collects all files from $c_\mathrm{old}$
that contain at least one test function for test localization.
It presents the list of files and the issue description
$d_\mathrm{issue}$ to the LLM and asks it to generate 10
relevant test file names. Our initial findings on the dev set of SWE-bench
indicated that the Top-10, Top-5, and Top-1 accuracy for test file localization
with GPT-4o are 83.6\%, 76.0\%, and 59.1\%, respectively.
We restricted file retrieval to 10 files so as not to overwhelm the
contexts for subsequent LLM calls.
Next, the localizer validates the file names by comparing them with
the previously collected file list and drops the ones that do not match.
After localizing test files, it makes a second LLM call with
file names and test function names from those files. The model chooses the
test files and functions relevant to~$d_\mathrm{issue}$.
The localizer again validates
the retrieved file names, but instead of dropping any hallucinated file names,
replaces them with the file names with minimal edit distance. This ensures
validity of identified files, which is essential for the subsequent
test-generation phase.
Note that even if the localizer chooses the wrong test file, if the
file is at least valid, the test generator may still succeed.
Focal file and function localization follows a similar two-LLM-call approach.
Figures~\ref{fig:focal1} and~\ref{fig:focal11} in the Appendix illustrate the LLM prompts
for localization.

\paragraph{Self-Reflective Action Planner.}

The second phase of \solx creates a \emph{plan}, which is a list of
actions for generating the fail-to-pass tests~$y$.
There are three kinds of actions: read, write, or modify.
A \emph{read}~$f$ action reads a function~$f$ from $c_\mathrm{old}$ to
use as context in a prompt.
A \emph{write}~$f$ or \emph{modify}~$f$ action declares the intent
to write a new test function or modify an existing test function.
Here, $f$ is a file and function name (in the planner, write and
modify actions do not yet include the exact code for the test
function, which is left to the test generator phase of \solx).
The planner starts by executing read actions for the
files and functions provided by the localizer.
Next, it prompts an LLM with the function definitions, the issue
description $d_\mathrm{issue}$, and instructions to make an initial
plan, restricted to only read actions.
The next step of the planner validates the planned actions: it checks
whether the actions generated by the LLM refer to valid file and
function names.
The final step of the planner is ``reflect and improve plan'', an LLM
call with a prompt including feedback from validation.
At this point, the plan is no longer restricted to only read actions,
and can also contain a write or modify action. Section~\ref{appendix_action_counts} in the appendix provides statistics on read/write/modify actions generated by the planner.
The model is also instructed to self-reflect on the proposed plan with
one of three possible outcomes: ``Satisfied'', ``Unsatisfied'', and
``Unsure''.
If the model chooses ``Satisfied'', \solx moves forward to the
test-generation phase.
For other options, it returns to the validation step and then repeats
the ``reflect and improve plan'' step.
This process is repeated at most five times.
In most cases, the model is satisfied with the plan in the first two
turns.
The two planner prompts are presented in Figures~\ref{plan1} and~\ref{plan2} (Appendix).

\paragraph{Test Generator.}

The test-generation phase of \solx executes the actions computed by the planner,
which can involve generating a new test (write action) or updating an existing
test (modify action). To guide
the LLM in this task, we extract the test structure and imports from the
localized test file and make them available in the prompt.
This reduces the burden on the LLM to generate imports.
The file structure is relevant for new tests, to
determine their insertion point in the test file.
\solx uses a different prompt for a write vs.\ a modify action,
illustrated in Figures~\ref{modifyprompt} and~\ref{fig:writeprompt} in the Appendix.
%% As we have plans from the self-reflective action planner phase, it's time to
%% execute them. We execute all of the planned actions. We also collect the test
%% structure and imports from the test file. The exposure to imports will prevent
%% the model from regenerating them, thus reducing some burden. The structure is
%% important, especially for new tests, to find where to insert them in the
%% file. Note that this structure is not important for function modification
%% because we already know where to insert the function. We have two separate
%% prompts for modified and new tests.
For new tests, the model needs to generate the preceding function name in
addition to the test.
Unlike most SWE agents, \solx does not try to generate diffs; instead,
it asks the model to generate complete test
functions, even for the modification case.
Since model training data tends to contain more complete functions
than diffs, we expect the model to perform better at generating
functions.

To handle missing or hallucinated imports, \solx includes an import-fixing step
in this phase, where it looks at model-generated imports and
%% The LLM may forget or hallucinate the necessary imports. We have an import
%% fixing phase where we look into two sources of information: the model-generated
%% import
linting errors detected using
Flake8\footnote{\url{https://flake8.pycqa.org/en/latest/}} (a static analysis tool)
to identify missing imports. Note that Flake8 reports different styling errors;
we manually curated the error codes to catch name-related errors.  In case of
missing imports, we add a dummy import to the function. Then, we take the
model-generated and dummy imports and try to find the imported module among the
files in the codebase. If we find the module, we replace the
model-generated/dummy import with the one from the codebase; otherwise, we
continue with the model-generated/dummy import as a fallback. Finally, we add
the function in the codebase and generate a git diff to create the test
patch.

\subsection{\soly: Ensemble using Multi-Sampling with Heterogeneous Prompting}

\solx has several components (e.g., localizer, planner), but they are not
perfect.  The file localizer accuracy for focal and test localization is 82.4\%
and 70.6\% with GPT-4o, respectively.  As the output from localization serves as
input to later LLM calls, those later calls may be affected by inaccuracies in
localization.  Conversely, LLMs can sometimes generate fail-to-pass tests in a
zero-shot setup, even without any context.  So, selectively including and
excluding parts of localization may produce different fail-to-pass tests that
are not generated by \solx.  \soly uses the test generated by \solx (T1) and
adds four new tests (T2--T5), obtained by skipping the planner stage, and
including neither, one, or both of focal and test localization.  In other words,
\soly runs the test generator stage five times with different,
\emph{heterogeneous} prompts.  We favor heterogeneous prompts with greedy
decoding over homogeneous prompts with higher model temperature because our
initial experiments showed that the latter yields poorer tests and lower
diversity.

To pick a single test among the five candidate tests, we run the five tests on
$c_\textrm{old}$ (recall that $c_\textrm{new}$ is not available yet) and analyze
the execution logs.  If a test passes on $c_\textrm{old}$, we just discard that
test because it violates the fail-to-pass criterion.  We classify the remaining
tests into three groups: assertion failure, other failure~(the test runs but
produces the wrong output), and error~(the test does not run properly, e.g.,
because of wrong syntax or an exception in a fixture).  We pick a test from the
first non-empty group to maximize the chance that it failed for the right reason
(i.e., it reproduces the bug described in the issue).  If the selected group has
multiple tests, we break the tie with a pre-determined ordering of the five
prompts that favors tests from prompts with more or better information: \solx,
followed by the prompts with both localizers, test localizer only, focal
localizer only, and neither localizer, in that order.

\subsection{Baseline: Zero-shot Test File Generation}

Recent instruction-tuned LLMs excel at following
instructions~\cite{peng2023instruction,zhang2023instruction}. We propose a
simple zero-shot approach to generate a fail-to-pass test given the repository
name and the issue description. % (see detailed prompt in Figure~\hl{X}). 
Given the prompt, the model generates a complete test file with all necessary imports to
make it compilable. In real scenarios, test files usually have multiple test
cases, but this baseline usually generates only a single test per file. 

% Commented this part as it's minor implementation detail
% after discussion with Toufique
% The
% generated test file (we call it \texttt{\small test\_tdd.py}) needs to be placed
% in the right directory for the imports in the test to work. Fortunately, all
% Python projects in \tdd have at least one directory called \texttt{\small
%   tests}. %; some projects have multiple such directories.  So we follow the
% simple approach of searching for the \texttt{\small tests} directory and placing
% \texttt{\small test\_tdd.py} in that directory.
% After that, we execute git diff to have access to the test patch.

%% file: benchmark.tex
\begin{figure*}[h]
    \vspace{-.1cm}
    \centering
    \includegraphics[width=.9\textwidth]{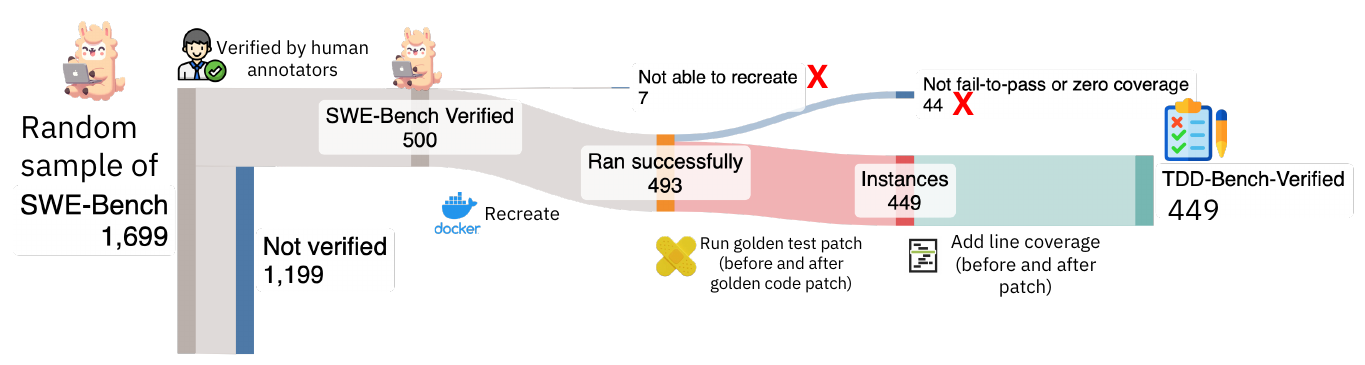}
    \vspace{-.4cm}
    \caption{Overall flow of TDD-bench dataset filtering starting from SWE-bench verified.}
    \label{fig:tdd-flow}
    \vspace{-.3cm}
\end{figure*}

\section{\tdd Benchmark}
\label{sec:tddbench}

This section presents \tdd, a new benchmark that supports evaluation of
techniques for generating tests from an issue description and an old code
version, without access to new code that resolves the
issue~(see problem statement in Section~\ref{sec:problem}).

\subsection{\tdd Evaluation Harness}\label{sec:eval_harness}

\begin{figure}[t]
  \centerline{\includegraphics[width=.85\columnwidth]{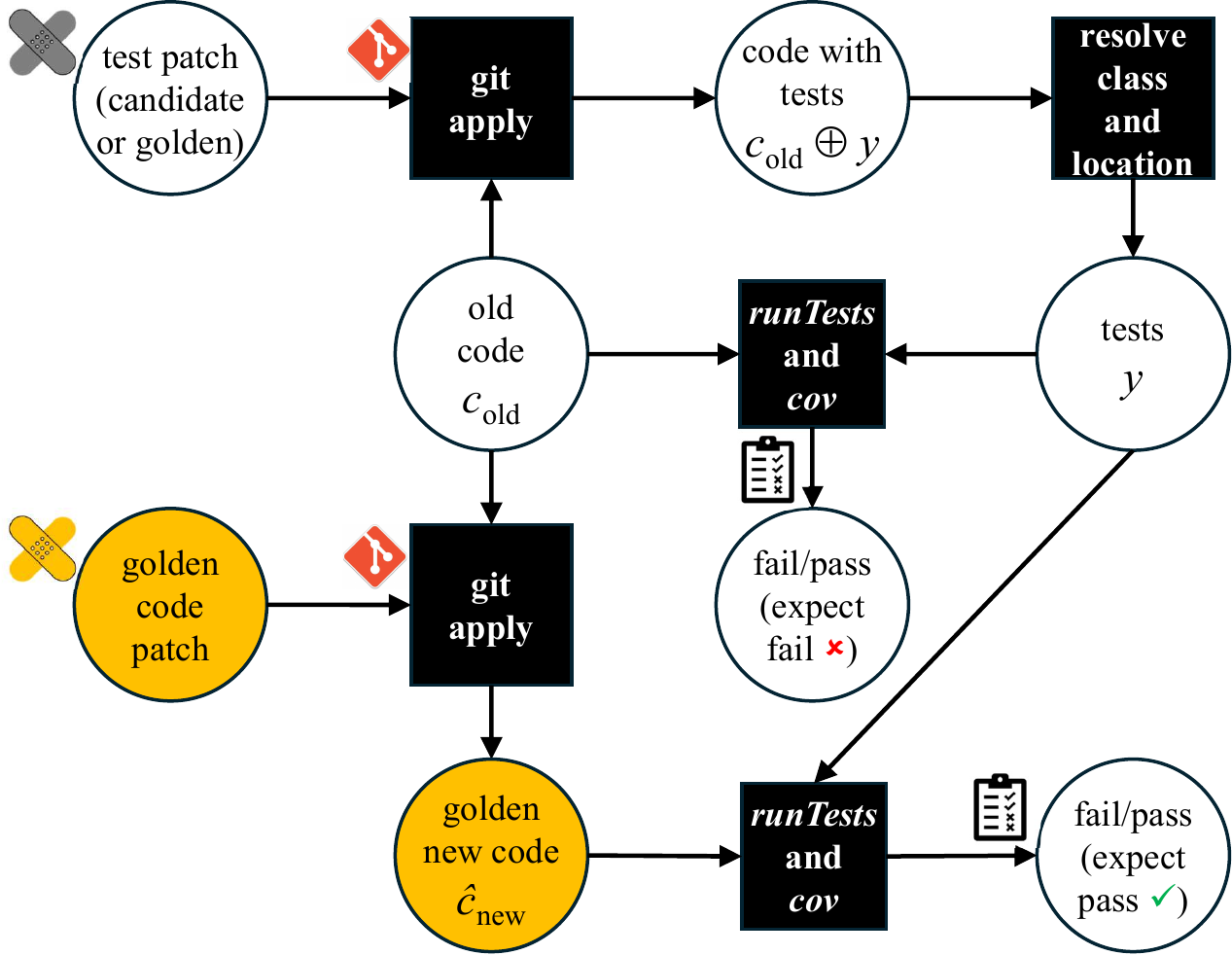}}
  \vspace{-.3cm}
  \caption{\label{fig:eval_harness}Evaluation harness for \tdd.}
  \vspace{-.3cm}
\end{figure}

\cref{fig:eval_harness} shows the harness for evaluating tests $y$, which
typically come from a $\mathit{genTests}$ solution, but the harness can also be
applied on golden tests $\hat{y}$ mined from a pull request~(PR).  The
evaluation harness runs in a containerized environment.  Starting at the top
left, tests come in the form of a patch, which is applied (via \texttt{\small
  git\,apply}) on the old code~$c_\mathrm{old}$.  Next, the harness analyzes the
resulting code $c_\mathrm{old}\oplus y$ to resolve the contributed test
functions~$y$.  The harness then executes $y$ while avoiding running other tests
that occur in the same file but were not contributed in the patch.  This yields
test results, including coverage achieved on the old code.  At least one of
these tests should fail, indicating that the tests reproduce the
issue at hand.

Moving to the bottom part of \cref{fig:eval_harness}, the code changes come from
the golden code patch mined from the same PR, which is applied on
$c_\mathrm{old}$ to obtain the golden new code $\hat{c}_\mathrm{new}$.  The harness
executes the tests $y$ again, this time on the new code, to obtain a second set
of test results.  This time, all tests should pass, to validate that the issue
was indeed resolved. An example test patch is presented
in~\cref{fig:test_patch}.

\subsection{Dataset Filters}\label{sec:tdd_filters}

\tdd builds upon prior work from SWE-bench~\cite{jimenezswe} and SWE-Bench Verified~\cite{chowdhury_et_al_2024}.  SWE-bench uses filters to keep
only those mined instances $x$ for which the set of golden tests $\hat{y}$ contains at least some tests
that fail on $c_\mathrm{old}$ and pass on the golden new code
$\hat{c}_\mathrm{new}$ from the same PR.  SWE-Bench Verified is a subset of
SWE-Bench, consisting of 500 instances further vetted by human
annotators~\cite{chowdhury_et_al_2024}.  The annotators filtered out instances
where the issue description $d_\mathrm{issue}$ was underspecified or the golden
tests $\hat{y}$ were overly specific, i.e., would reject some valid new
code~$c_\mathrm{new}$.  They also removed some instances where tests failed due
to environment problems. %instead of the solution.

In the same spirit, \tdd applies more filters to obtain an even higher-quality
dataset.  In a nutshell, the filtering process applies the \tdd evaluation
harness (\cref{fig:eval_harness}) to the golden tests $\hat{y}$ from
the original PR.  Specifically, substituting $\hat{y}$ wherever $y$ occurs in
\cref{fig:eval_harness} checks whether the PR indeed contributed tests that went
from failing to passing.  We filter out any instance where the contributed tests
do not satisfy that criterion.  Although the human annotators of SWE-bench
Verified were diligent, a few instances slipped past their filters, and we drop
those for \tdd.

\cref{fig:tdd-flow} illustrates the filtering process.  Starting from the 500 instances of SWE-bench Verified,
we first drop 7 instances whose environment we could not
recreate. Next, we run the test harness on the golden tests
$\hat{y}$. This filters out 44 additional instances because the tests
do not have the expected fail-to-pass behavior (25 instances) or have zero
line coverage on the golden code patch (19 instances).
In the end, 449 high-quality instances remain across 12 repositories.
\cref{tbl:tdd-stat} summarizes key statistics of \tdd.

% Starting with the 500
% instances of SWE-bench Verified, we first drop 7 instances whose environment we
% could not recreate.  Next, we run the test harness on the golden tests
% $\hat{y}$.  This filters out 44 additional instances because the tests do not
% have the expected fail-to-pass behavior or have zero line coverage on the golden
% code patch. 

\subsection{Evaluation Metric}
\label{subsec:eval_metric}

Passing a test does not necessarily mean a patch is adequate to address the
issue.  \citet{aleithan2024swe} reported that 31.1\% of the passed code patches
in SWE-Bench are suspicious due to weak test cases. To evaluate test adequacy,
we compute the coverage of the submitted test patch.  One key
difference between SWE-Bench Verified and \tdd is that the former runs
an entire test file to evaluate the submitted patch, whereas we
only run the contributed tests $y$.
Not running other test cases enables us to precisely compute coverage of~$y$.
If the tests are relevant, they should cover the deleted
lines in $c_\mathrm{old}$ and the added lines in $\hat{c}_\mathrm{new}$.  We
integrated the Python Coverage package into the 12 repositories and updated the
test scripts to allow us to run specific test cases and compute coverage for them.

We define the $\mathit{tddScore}$ metric that evaluates the quality of tests
generated by a solution $\mathit{genTests}$ over a set
\mbox{$X=\{x_0,x_1,\ldots\}$} of instances.  It returns a number between 0 and
100, the higher the better.  It is defined as 100 times the arithmetic mean of
the per-instance scores:
{\small
\[\mathit{tddScore}(X, \mathit{genTests})
  = \frac{100}{|X|}\sum_{x\in X} \mathit{tddScore}\big(x, \mathit{genTests}(x)\big)
\]}
Given a set of tests $y=\mathit{genTests}(x)$ submitted for an
instance, the per-instance score is a product of two factors:
\[\mathit{tddScore}(x, y)
  = \mathit{failToPass}(x, y) \cdot \mathit{adequacy}(x, y)
\]
The first factor is a binary correctness metric, using the indicator function
for the tests $y$ failing on the old code times the indicator function for the
tests $y$ passing on the new code.  While the solution $\mathit{genTests}$ only
has access to the old \mbox{code $c_\mathrm{old}$}, the evaluation metric also
uses the hidden golden new \mbox{code $\hat{c}_\mathrm{new}$} right after the
issue was fixed.
{\small
\[\begin{array}{@{}l@{}}\mathit{failToPass}(x, y) =\\
  \;\;I\big(\mathit{fail} \in \mathit{runTests}(y, c_\mathrm{old})\big)
  \cdot I\big(\mathit{fail} \notin \mathit{runTests}(y, \hat{c}_\mathrm{new})\big)
\end{array}\]}
The second factor is a fraction between 0 and 1 based on test coverage on
the old and new code:
{\small
\[\begin{array}{@{}l@{}}\mathit{adequacy}(x, y) =\\
  \displaystyle\;\;
    \frac{  |\mathit{cov}(y, c_\mathrm{old}) \cap (c_\mathrm{old}\setminus\hat{c}_\mathrm{new})|
          + |\mathit{cov}(y, \hat{c}_\mathrm{new}) \cap (\hat{c}_\mathrm{new}\setminus c_\mathrm{old})|}
         {  |c_\mathrm{old}\setminus\hat{c}_\mathrm{new}| \quad
          + \quad\; |\hat{c}_\mathrm{new}\setminus c_\mathrm{old}|}
\end{array}\]}
Adequacy focuses on just the coverage of lines added and deleted when going from
the old code to the new code, because those are the most relevant lines to be
tested.  In the above, $\mathit{cov}(y,c)$ is the set of lines covered by
running tests $y$ on code~$c$; $(c_\mathrm{old}\setminus\hat{c}_\mathrm{new})$
is the set of lines deleted by the PR patch; and $(\hat{c}_\mathrm{new}\setminus
c_\mathrm{old})$ is the set of lines added by the PR patch. 
We evaluate adequacy jointly for added and deleted lines, as some code patches may contain only added or deleted
lines.

% We had initially
% considered defining adequacy with two separate fractions for the deleted
% and added lines.  However, that was not only poorly weighted but brittle,
% because in some cases, the numerator or denominator of one of the fractions was
% zero.

%% file: result.tex
\section{Evaluation}

We conducted an extensive set of empirical studies, evaluating the effectiveness
of our approach (\S\ref{sec:rq1}) and the components of \solx and \soly
(\S\ref{sec:rq2}), comparing our approach against existing techniques
(\S\ref{sec:rq3}), and investigating the cost effectiveness of \solx
(\S\ref{sec:rq4}), characteristics of the generated tests (\S\ref{sec:rq5}),
usefulness of the generated tests in supporting automated program repair
(\S\ref{sec:rq6}), and possible effects of data contamination (\S\ref{sec:rq7}).

\subsection{Experiment Setup}
\label{sec:setup}

The evaluation used the closed-source GPT-4o~(gpt-4o-2024-08-06)
and the open-source Mistral-large model~(123 billion parameters).
All experiments used greedy decoding. For each instance, \solx makes 7--11 LLM
calls for T1. \soly makes one additional call for each of the other four tests (T2-T5) after the
localization stage. To evaluate using the generated tests for SWE agents, we conducted a large-scale
experiment with 22 systems from the SWE-Bench leaderboard. We ran \mbox{22 × 449 × 5 =
49,390 Docker} containers or tests (one Docker container per test) to report the
results.

\subsection{Effectiveness of \solx, \soly, and the Baseline}
\label{sec:rq1}

\cref{tbl:otter} shows that GPT-4o-based \solx and \soly perform well on test
generation, creating fail-to-pass tests for 31.4\% and 37\% of the instances,
respectively, whereas the baseline produced such tests for 18.7\% of the
instances. The improvements are also reflected in $\mathit{tddScore}$. We
observe similar performance improvements with Mistral-large. The pass@5 rate
(where one of the five tests is fail-to-pass) for \soly is 44\% for GPT-4o and 37\%
for Mistral-large.

\begin{table}[t]
\centering
\caption{Performance of \solx, \soly, and baseline technique on TDD-Bench-Verified.}
\vskip 0.05in
\resizebox{\columnwidth}{!}{%
\renewcommand{\arraystretch}{1.2}% Tighter
\begin{tabular}{@{}llrrrr@{}}
\toprule
\multicolumn{1}{c}{Model}      & \multicolumn{1}{c}{Approach} & \multicolumn{1}{c}{\begin{tabular}[c]{@{}c@{}}\# of fail-to-pass\\ test\end{tabular}} & \multicolumn{1}{c}{\begin{tabular}[c]{@{}c@{}}\# of fail-to-pass \\ test in (\%)\end{tabular}} & \multicolumn{1}{c}{tddScore} & \multicolumn{1}{c}{Coverage}  \\ \midrule
\multirow{3}{*}{Mistral-Large} & Zero-shot                    & 57                                                                                    & 12.7                                                                                          & 11.8                         & 60.6                         \\
                               & \solx                       & 121                                                                                   & 26.9                                                                                           & 25.1                         & 70.5                         \\
                               & \soly                      & 144                                                                                   & 32.1                                                                                           & 28.6                         & 70.4                         \\ \midrule
                             
\multirow{3}{*}{GPT4o}         & Zero-shot                    & 84                                                                                   & 18.7                                                                                             & 17.2                        & 60.0                         \\
                               & \solx                        & 141                                                                                   & 31.4                                                                                           & 29.4                         & 70.6                         \\
                               & \soly                     & 166                                                                                   & 37.0                                                                                             & 32.4                         & 71.5                        \\  \bottomrule

\end{tabular}}
\label{tbl:otter}
\vspace{-15pt}
\end{table}

\subsection{Ablation Study}
\label{sec:rq2}

\cref{tbl:ablation} shows the ablation study for \solx (T1). We also present the
individual performance for the other four tests~(T2--T5) produced by \soly. We
can see that all the components contribute to \solx's performance. Without
action planning, we lose more than 14\%--20\% of fail-to-pass tests for GPT-4o
and 21\%--36\% of the tests for Mistral-large.

\begin{table}[t]
\centering
\caption{Contribution of each component of \solx.}
\vskip 0.05in
\resizebox{\columnwidth}{!}{%
\renewcommand{\arraystretch}{1.2}% Tighter
\begin{tabular}{@{}lllrrr@{}}
  \toprule
  \multicolumn{1}{c}{Model}      & \multicolumn{1}{c}{Component}   & \multicolumn{1}{c}{Approach}              & \multicolumn{1}{c}{\begin{tabular}[c]{@{}c@{}}\# of fail-\\to-pass\end{tabular}} & \multicolumn{1}{c}{tddScore} & \multicolumn{1}{c@{}}{\begin{tabular}[c]{@{}c@{}}Change in  \\ tddScore\%\end{tabular}} \\ \midrule
  \multirow{9}{*}{\begin{tabular}[c]{@{}c@{}}Mistral-\\large\end{tabular}} & -                              & \solx (T1)                                & 121                                    & 25.7                         & -                                                                                   \\
                                 & \multirow{3}{*}{Action Planner} & without Action Planning (complete)* (T2)  & 96                                     & 20.2                         & -21.4                                                                                \\
                                 &                                 & without Plan Refinement (Just 1 attempt)  & 115                                    & 23.8                         & -7.4                                                                                 \\
                                 &                                 & without Action Validation                 & 107                                    & 22.1                         & -14.0                                                                                  \\
                                 & \multirow{3}{*}{Localizers}     & without Focal Localization* (T3)          & 96                                     & 20.2                         & -21.4                                                                                \\
                                 &                                 & without Test Localization* (T4)           & 77                                     & 16.5                         & -35.8                                                                                \\
                                 &                                 & without  Focal \& Test Localization* (T5) & 81                                     & 17.2                         & -33.1                                                                                \\
                                 & \multirow{2}{*}{Test Generator} & without Fixing Import                     & 117                                    & 24.6                         & -4.3                                                                                 \\
                                 &                                 & without Imports at Generation             & 114                                    & 23.7                         & -7.8                                                                                 \\ \midrule
  \multirow{9}{*}{GPT-4o}        & -                              & \solx (T1)                                & 141                                    & 29.4                         & -                                                                                   \\
                                 & \multirow{3}{*}{Action Planner} & without Action Planning (complete)* (T2)  & 110                                    & 23.6                         & -19.7                                                                                \\
                                 &                                 & without Plan Refinement (Just 1 attempt)  & 130                                    & 27.5                         & -6.5                                                                                 \\
                                 &                                 & without Action Validation                 & 120                                    & 25.7                         & -12.6                                                                                \\
                                 & \multirow{3}{*}{Localizers}     & without Focal Localization* (T3)          & 115                                    & 25.2                         & -14.3                                                                                \\
                                 &                                 & without Test Localization* (T4)           & 107                                    & 24.2                         & -17.7                                                                                \\
                                 &                                 & without  Focal \& Test Localization* (T5) & 110                                    & 24.2                         & -17.7                                                                                \\
                                 & \multirow{2}{*}{Test Generator} & without Fixing Import                     & 128                                    & 26.7                         & -9.2                                                                                 \\
                                 &                                 & without Imports at Generation             & 130                                    & 26.7                         & -9.2  \\ \bottomrule      
                                 \multicolumn{5}{l}{* is not followed by action planning}                                                                         
  \end{tabular}}
\label{tbl:ablation}
\vspace{-12pt}
\end{table}

%\multicolumn{4}{l}{* is not followed by action planning}  

\subsection{Comparison with the Approaches of M{\"u}ndler \etal}
\label{sec:rq3}

\citet{mundler2024swtbench} proposed a set of approaches for
generating fail-to-pass tests. We ran \solx and \soly on their dataset to study
how the approaches compare. They also evaluated zero-shot approaches, which
differ from our zero-shot baseline.
All of their approaches (including the zero-shot ones) instruct the
model to generate a novel code diff format introduced by their paper.
Two of their approaches use a proposed patch
in the prompt. M{\"u}ndler \etal's SWE-agent and SWE-agent+ approaches are
derived from SWE-Agent, which was originally designed for generating code
patches~\cite{sweagent2}. \cref{tbl:swt} shows the results. \solx and \soly
perform better than their best-performing approach, generating 70 (25.4\%) and
80 (29.0\%) fail-to-pass tests compared to 53 (19.2\%) fail-to-pass tests
generated by SWE-agent+.

\soly uses execution logs on the current code base $c_\mathrm{old}$ which give it an advantage 
in the final selection stage. The feedback works as a contributing factor to the superior performance of \soly.
Note that \solx does not use any execution logs, yet it still performs significantly better than LIBRO (25.4\% vs.\ 15.2\% in \cref{tbl:swt}).
The novel code diff format proposed by M{\"u}ndler \etal has similarities to our approach. 
However, their format requires the model to perform additional tasks such as writing the file name, change type, and line number in response to one LLM call. 
Additionally, to instruct the model to generate a specific format, the authors had to include an example that is not relevant to the issue itself.
In our generation step, the model only needs to generate the test (and prior function name for positioning new tests).

\begin{table}[t]
\centering
\caption{Comparing with approaches proposed by \citet{mundler2024swtbench} on the 276 instances of their SWT-Lite.}
%(not \tdd).}
\vskip 0.05in
\resizebox{.7\columnwidth}{!}{%
\renewcommand{\arraystretch}{1.2}% Tighter
\begin{tabular}{lrr}
\toprule
\multicolumn{1}{c}{Approach} & \multicolumn{1}{c}{\begin{tabular}[c]{@{}c@{}}\# of Fail-to-pass \\ Tests\end{tabular}} & \multicolumn{1}{c}{in (\%)} \\ \midrule
ZeroShot (GPT-4)             & 16                                                                                      & 5.8                        \\
ZeroShotPlus* (GPT-4)        & 28                                                                                      & 10.1                        \\
LIBRO* (GPT-4)               & 42                                                                                      & 15.2                         \\
AutoCodeRover (GPT-4)        & 25                                                                                      & 9.1                         \\
SWE-Agent (GPT-4)            & 46                                                                                      & 16.7                         \\
SWE-Agent+ (GPT-4)           & 53                                                                                      & 19.2                       \\ \midrule
\textbf{\solx} (GPT-4o)      & \textbf{70}                                                                             & \textbf{25.4}  \\
\textbf{\soly}  (GPT-4o)     & \textbf{80}                                                                             & \textbf{29.0}  \\  \bottomrule
%\textbf{\soly (pass@5)}                      & \textbf{105}                                                                                         & \textbf{38.0}  \\ \bottomrule
\multicolumn{3}{l}{* uses ``proposed patch'' while generating tests} 
\end{tabular}
}
\label{tbl:swt}
\vspace{-15pt}
\end{table}

\subsection{Cost Effectiveness of \solx}
\label{sec:rq4}

\begin{figure}[!ht]
    \centering
    \includegraphics[width=.8\columnwidth]{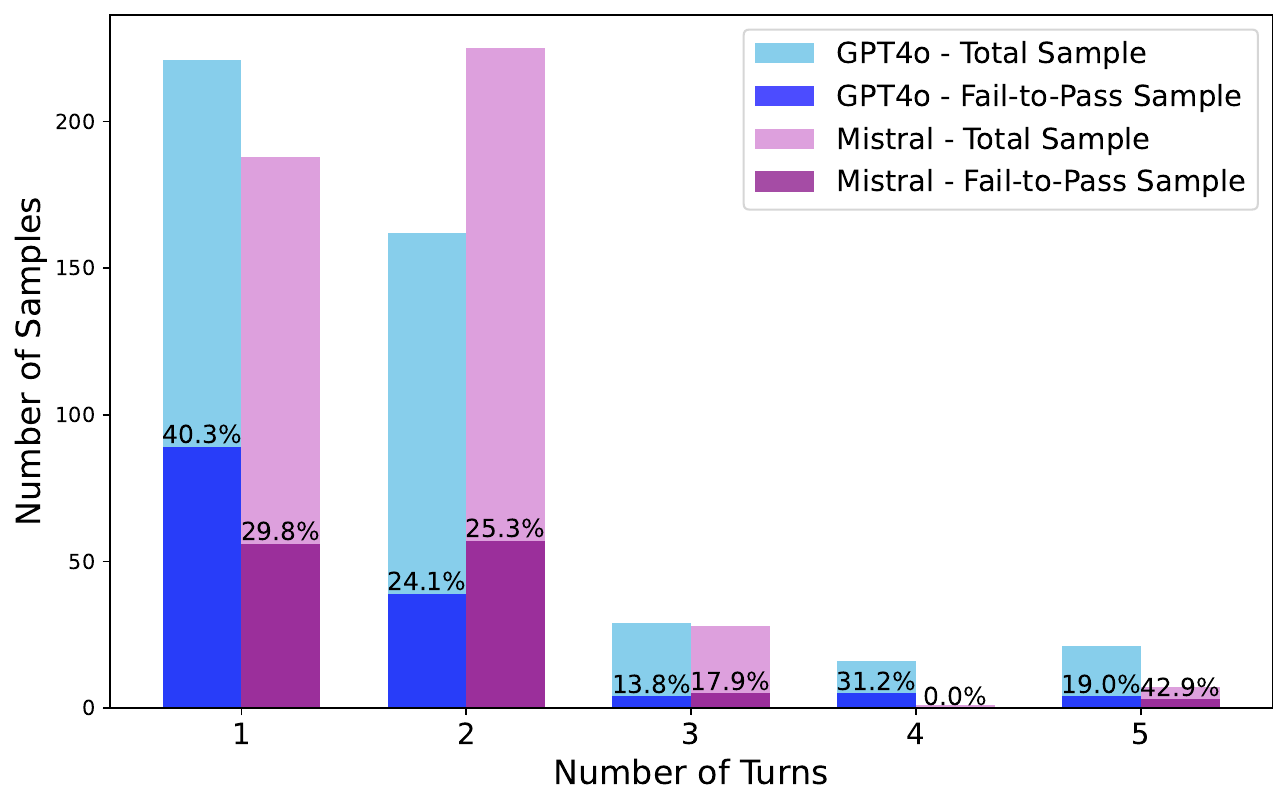}
    \vspace{-10pt}
    \caption{Number of turns taken in ``reflect and improve plan'' step.}
    \label{fig:turn}
    \vspace{-15pt}
\end{figure}

\cref{tbl:cost} presents the cost for invoking the GPT-4o model to process 449
instances with \solx. Each sample requires an average of \$0.06.
%Note that apart from the ``Reflect and Improve the Plan'' step, we make a total of 6 calls. 
The ``reflect and improve plan'' step can make 1--5 LLM
calls. \cref{fig:turn} shows that, for more than 80\% of the instances, both
GPT-4o and Mistral are satisfied within two calls. Therefore, the total calls
vary from 7--8 for most instances. The cost is very low because we do not
accumulate context from prior calls in subsequent calls.
%However, the models generate fail-to-pass tests even with more than 2 turns at the ``reflect and Improve the plan'' steps. 
We do not discuss the cost for Mistral-large because the model was
hosted locally.  \soly makes four additional calls and reuses the output from
localizers. The total cost for \soly~(which includes \solx) is \$0.09 per instance.

\begin{table}[t]
\centering
\caption{Cost for running \solx and \soly with GPT-4o.}
\vskip 0.05in
\resizebox{.7\columnwidth}{!}{%
\renewcommand{\arraystretch}{1.2}% Tighter
\begin{tabular}{lrr}
\toprule
\multicolumn{1}{c}{Component} & \multicolumn{1}{c}{Cost} & \multicolumn{1}{c}{Cost/Sample} \\ \midrule
Focal Localization            & \$8.61                   & \$0.02                          \\
Test Localization             & \$9.63                   & \$0.02                          \\
Action + Generate             & \$10.94                  & \$0.02                          \\
\textbf{Total for \solx}               & \textbf{\$29.18}                  & \textbf{\$0.06}                          \\ \midrule
Additional Tests (T2-T5)      & \$11.20                  & \$0.02                          \\
\textbf{Total for \soly}             & \textbf{\$40.38}                  & \textbf{\$0.09}           \\ \bottomrule               
\end{tabular}}
\label{tbl:cost}
\vspace{-10pt}
\end{table}

\subsection{Characteristics of the Generated Tests}
\label{sec:rq5}

\paragraph{Coverage of the generated tests.}

We compared the coverage achieved by the \solx-generates tests and the golden
tests written by developers. We observe that, for the fail-to-pass tests,
\solx-generated and golden tests have very similar and high coverage---more than
90\% with both models (\cref{tbl:coverage}). However, for the other tests, the
coverage is quite low for \solx. This indicates that tests with higher coverage
are more likely to be fail-to-pass tests.

\begin{table}[t]
\centering
\caption{Comparing the coverage of model-generated and developer-written golden tests.}
\vskip 0.05in
\resizebox{.7\columnwidth}{!}{%
\renewcommand{\arraystretch}{1.2}% Tighter
\begin{tabular}{llrr}
\toprule
\multicolumn{1}{c}{Model}      & \multicolumn{1}{c}{Is Fail-to-pass?} & \multicolumn{1}{c}{Otter} & \multicolumn{1}{c}{Golden test} \\ \midrule

\multirow{2}{*}{Mistral-large} & Yes                                  & 93.1                      & 95.5                            \\
                               & No                                   & 63.8                      & 93.4                 \\  \midrule
                               
 \multirow{2}{*}{GPT-4o}        & Yes                                  & 93.7                      & 95.6                            \\
                               & No                                   & 60.0                      & 93.3                            \\

                               \bottomrule           
\end{tabular}}
\label{tbl:coverage}
\vspace{-10pt}
\end{table}

\paragraph{Prompts complementarity.}

\begin{figure}[t]
    \centering
    \includegraphics[width=.7\columnwidth]{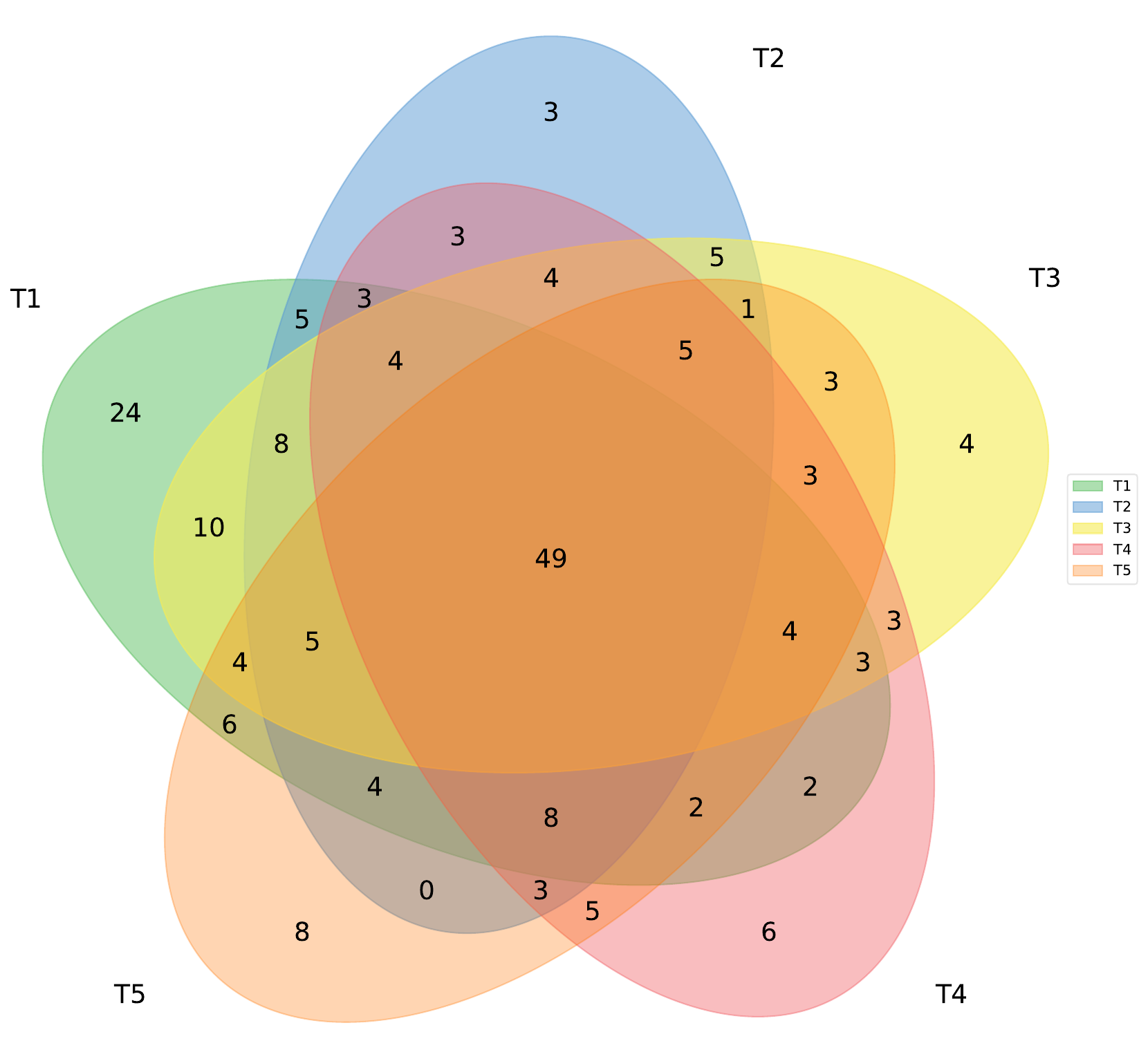}
    \vspace{-10pt}
    \caption{Number of instances with fail-to-pass tests generated by different prompts.}
    \vspace{-15pt}
    \label{fig:venn}
\end{figure}

\cref{fig:venn} presents the overlap among instances for which fail-to-pass
tests could be generated by different prompts using the GPT-4 model. Overall,
each of the prompts is successful on some instances on which the other prompts
fail, with T1 (\solx with GPT-4o) achieving the most success in this
respect---producing fail-to-pass tests for 24 instances on which none of the
other prompts succeeded in generating such tests. Thus, combining the results
from the different prompts increases the fail-to-pass rate to 44\% (38\% for
Mistral) at pass@5. This indicates the potential of incorporating different
prompts for test generation.

\paragraph{Detailed analysis of the tests.}
\cref{tbl:analysis} presents the analysis of the \solx-generated tests using GPT-4o model from different perspectives.
We see that the fail-to-pass rate is higher when patching an existing test than when adding a new test, which is expected because the model can have better context from the existing test.
Writing a new test is inherently more difficult than modifying an existing test. 
It is expected that the test will fail on the old codebase. However, we found 69 samples in \solx where the test passed on the old codebase.
Also, our analysis shows that tests with assertion failures have a higher success rate (57.1\% fail-to-pass rate) compared to other groups. 
We did not see much impact of focal localization on the performance, with 31.9\% and 29.1\% for correct and incorrect localization, respectively. 
Test localization has a significant impact on the performance (36.9\% vs.\ 18.2\% for correct and incorrect localization). 
In our ablation study, we also found that test localization is more important than focal localization. In~\cref{halucination}, we discuss the impact of hallucination replacer in our pipeline.

\begin{table}[t]
  \centering
  \caption{Analysis of the \solx-generated tests using GPT-4o model from different perspectives.}
  %(not \tdd).}
  \vskip 0.05in
  \resizebox{\columnwidth}{!}{%
  \renewcommand{\arraystretch}{1.2}% Tighter
  \begin{tabular}{@{}llrrr@{}}
    \toprule
    \multicolumn{1}{c}{Perspective}       & \multicolumn{1}{c}{Category} & \multicolumn{1}{c}{\# of Sample} & \multicolumn{1}{c}{\#of fail-to-pass} & \multicolumn{1}{c@{}}{fail-to-pass rate} \\ \midrule
    \multirow{2}{*}{Type of Test}         & PatchExisting                     & 122                              & 50                                    & 41.0                                       \\
                                          & AdditionOnly                          & 327                              & 91                                    & 27.8                                       \\ \midrule
    \multirow{4}{*}{Test on old Codebase} & Pass                         & 69                               & 0                                     & 0                                          \\
                                          & AssertionFail                & 170                              & 97                                    & 57.1                                       \\
                                          & Fail                         & 101                              & 40                                    & 39.6                                       \\
                                          & Error                        & 109                              & 4                                     & 3.7                                        \\ \midrule
    \multirow{2}{*}{Focal Localization}   & Correct                      & 370                              & 118                                   & 31.9                                       \\
                                          & Wrong                        & 79                               & 23                                    & 29.1                                       \\ \midrule
    \multirow{2}{*}{Test Localization}    & Correct                      & 317                              & 117                                   & 36.9                                       \\
                                          & Wrong                        & 132                              & 24                                    & 18.2  \\ \bottomrule                                    
    \end{tabular}
  }
  \label{tbl:analysis}
  \vspace{-10pt}
  \end{table}

\paragraph{Heterogenous prompts vs.\ temperature.}
\soly scaled well with samples up to 5, giving 0.5\%-5.6\% improvement (see~\cref{scaling}).
\soly uses heterogeneous prompting instead of higher temperature to generate multiple samples. We make multiple LLM calls in different stages 
and multi-sampling in each stage would exponentially increase the test counts. Therefore, to compare heterogenous prompting with temperature, we generated 
5 samples at high temperature (1.0) in the last LLM call in the Test Generator phase of \solx (our best solution). 
Though the average number of fail-to-pass test goes up (117.8 vs.\ 116.6), the Top-5 and Top-1 (using \soly's ranker) 
results remain lower (Top-5: 173 vs.\ 197 and Top-1: 146 vs.\ 166). That means heterogenous prompting boosts up the overall ensemble 
performance if we compare at the same number of samples. 
\cref{heterogeneous_prompt} discusses more details.

% \subsection{Test generation and Code Repair}
\subsection{Test Generation and SWE Agents}
\label{sec:rq6}

%\subsubsection{Filtering Code Fixes with Tests }

\begin{figure}[!t]
    \centering
    \includegraphics[width=\columnwidth]{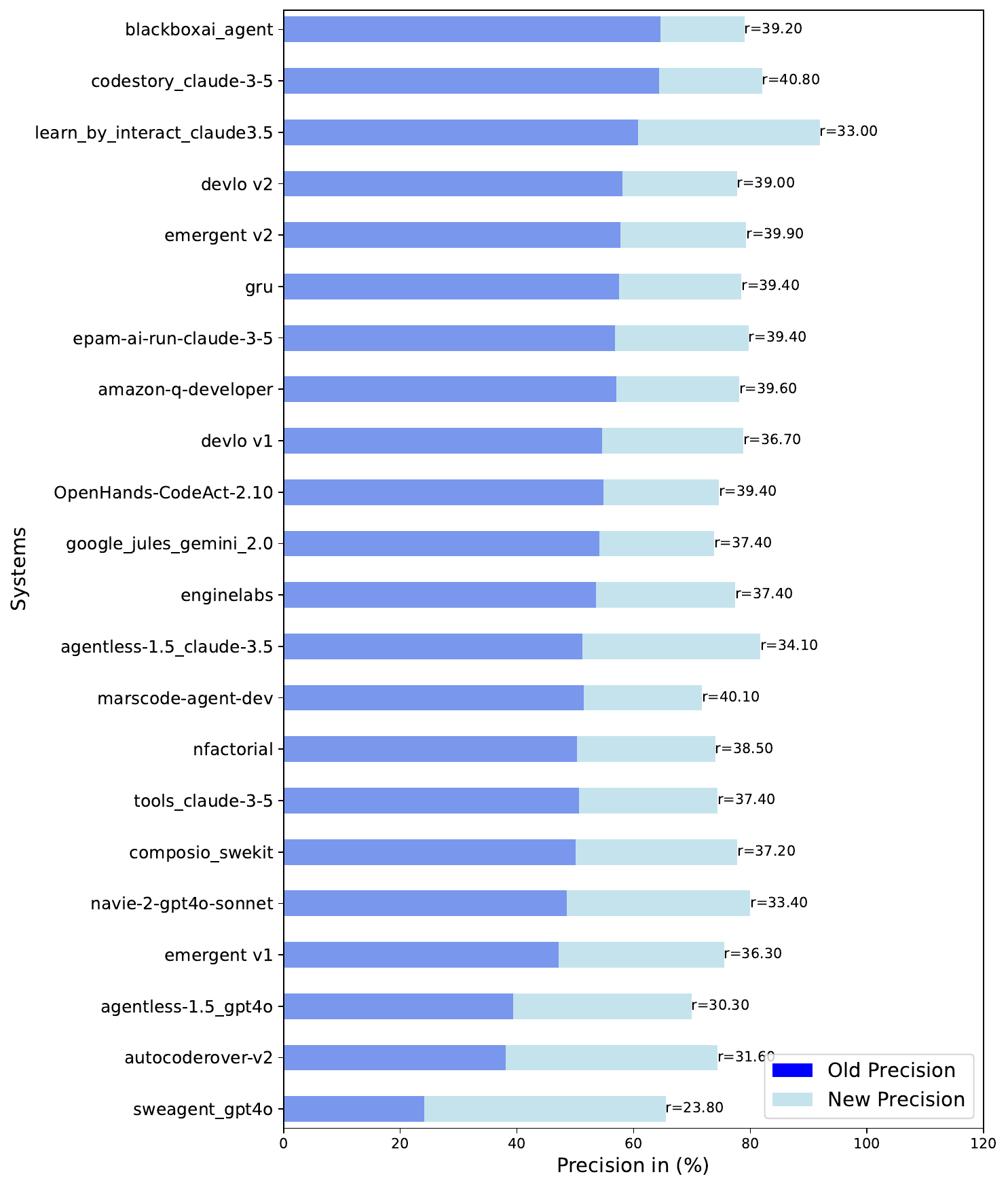}
    \vspace{-10pt}
    \caption{Precision of all 22 systems from the leaderboard. Recall is also mentioned at the top of each bar.}
    \vspace{-5pt}
    \label{fig:precision}
\end{figure}

The tests generated by our approach can be used for filtering bad code patches and
increasing the precision of solutions proposed by different systems from the
SWE-bench Verified leaderboard. We take the top 22 systems from the leaderboard
and run the five tests generated by \soly. We filter out a code patch if all the
tests fail on it. Figure~\ref{fig:precision} shows that this achieves a
precision of 65\% to 92\% while maintaining a decent recall of 30\%-41\%, except
for one system where the precision increased by 22\% to 167\%. Note that
\citet{mundler2024swtbench} achieved 47.8\% precision at 20\%
recall on SWE-Agent.
Using tests generated by \soly achieved much higher precision while maintaining
greater recall.

%\subsubsection{TBM: Ensemble on Lederboard}
%\subsubsection{Correlation between Test Generation and Code Repair}

Apart from filtering SWE-patches, we could use our tests to choose the best SWE-patch, which would be a good application of the \solx-generated tests. 
We have tried CodeT ranking on candidates from the top 3 leaderboard systems on SWE-Bench-Verified and observed 2\% improvement.
Note that some of these leaderboard solutions have already been through good rankers and used superior models. Improving upon these samples using a ranker may be difficult.

\subsection{Effect of Data Contamination}
\label{sec:rq7}

As TDD-Bench-Verified is based on historic GitHub issues, they may be included
in the pre-training data of the LLMs we use. To see whether the model simply
generates memorizes tests, we performed two different experiments.

%\subsubsection{Based on Model Data Cut-off Date} 
\paragraph{Model data cut-off date.}

\begin{figure}[t]
    \centering
    \includegraphics[width=.7\columnwidth]{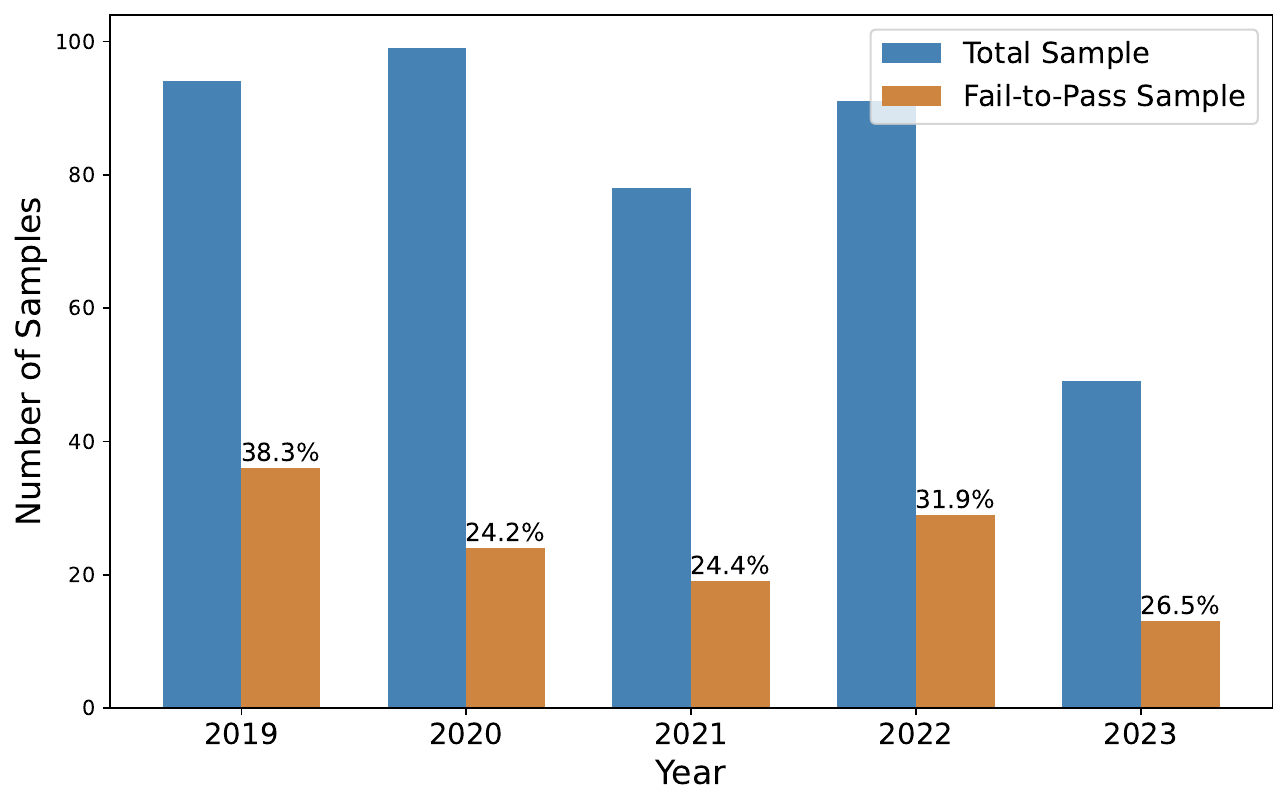}
    \vspace{-15pt}
    \caption{Sample distribution of total and fail-to-pass tests by year.}
    \vspace{-15pt}
    \label{fig:year}
\end{figure}

The cutoff date for the GPT-4o model is October 2023. Unfortunately, we have
only one sample dated post-cutoff and could not compare the two groups.  Popular
GitHub repositories evolve quickly. It is likely that only the snapshot taken
during the data collection process was seen by the model and that it performed
well on a specific year of data.  \cref{fig:year} shows the total and
fail-to-pass test distribution by year.  We did not observe any pattern in
performance among the distributions by year.  For example, for 2020, 2021, and
2023, we have very similar performance by the model.

%\subsubsection{Based on Test Similarity}
\paragraph{Test similarity.}

\begin{figure}[t]
    \centering
    \includegraphics[width=.9\columnwidth]{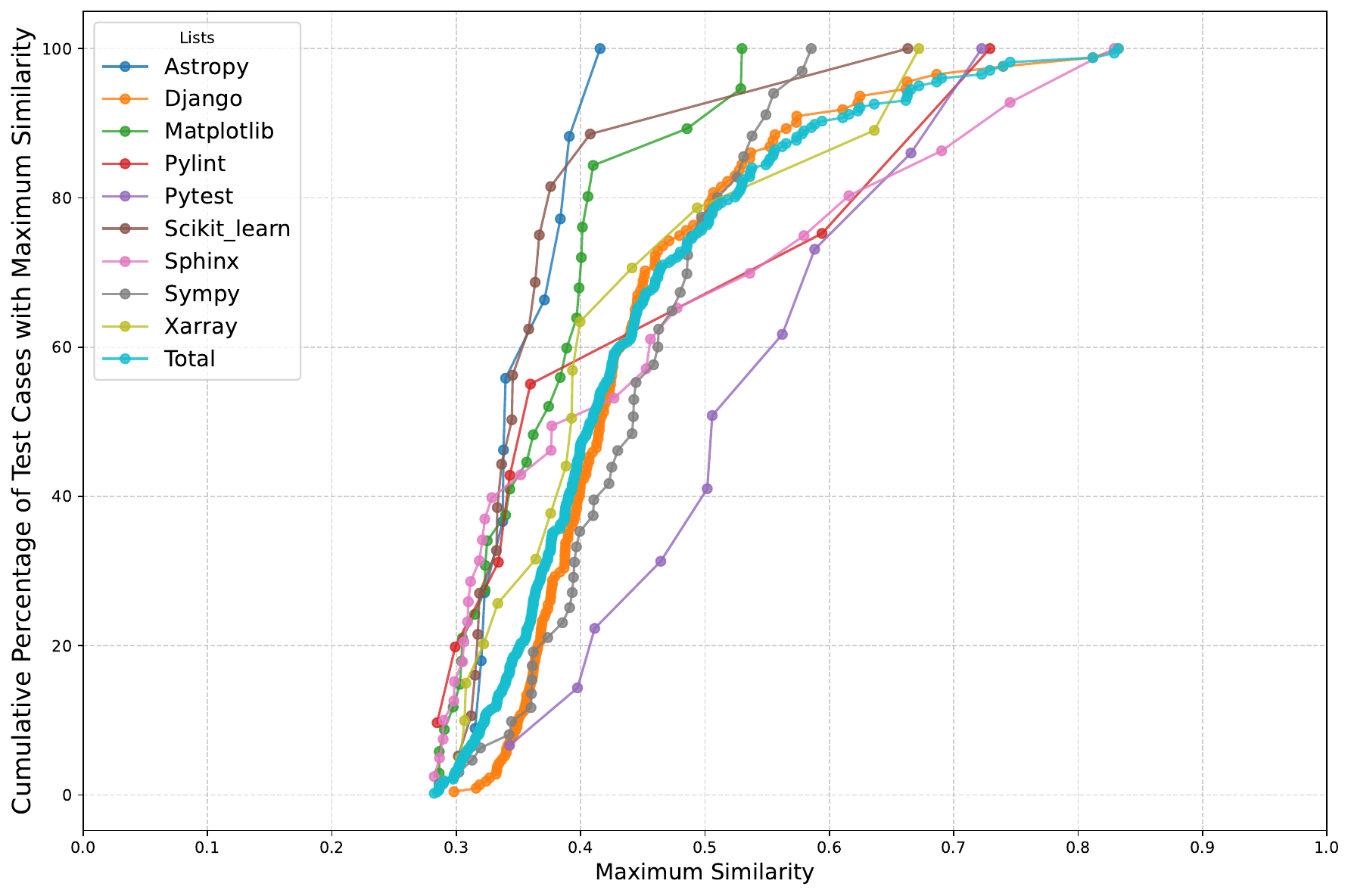}
    \vspace{-10pt}
    \caption{Cumulative percentage of \solx-generated new tests, using
      GPT-4o, with maximum similarity less than the similarity value
      shown on the x-axis.}
    \vspace{-15pt}
    \label{fig:write_contamination}
\end{figure}

We conducted another experiment following the approach of
\citet{schafer2024empirical}.
We compute the similarity score between the generated test from \solx
and the most similar test from the repository, as follows:
$\mathit{max}_{tp \in \mathit{TP}}(1-\frac{\mathit{dis}(t^{*},t_{p})}{\mathit{max}(\mathit{len}(t^{*}),\mathit{len}(t_{p}))})$, where $\mathit{TP}$ is
the set of test functions and $t^*$ is the generated test.
Figure~\ref{fig:write_contamination} shows the similarity scores for new
tests. For 90\% of the instances, the similarity score is less than 0.6.
Table~\ref{tbl:similarity_example} in the appendix shows some samples
with more than 0.5 similarity to give the reader some
idea. From our observation, even at similarity score of 0.7, the tests are
significantly different. As expected, for modified tests, the similarity is
higher. However, we did not observe any difference between fail-to-pass tests
and other tests (see appendix for figures). Therefore, the model is not simply generating memorized tests.

%\begin{enumerate}
%
%
%\item Coverage comparison between generated test patch \& golden test patch on the leaderboard
%\item Correlation of Test Generation and Code Repair
%\item Effect of Issue Description Length
%\item Method Complimentarity
%
%\item Precision increase
%\item Ensemble from the leaderboard
%
%\item data contamination *** Refer to swt-bench
%\end{enumerate}

%% file: limitation.tex
\section{Limitations}

One limitation of \tdd is that it is mined from 12 popular Python
repositories, so our findings may not apply to other programming
languages and repositories.  We note that SWE-bench, despite having
the same limitation, has been impactful, and one of the findings in
the SWE-bench paper is that ``difficulty does not correlate with issue
resolution date'', indicating that contamination problems (if any) are
minor~\cite{jimenezswe}.  Our results on data contamination
(\S\ref{sec:rq7}) indicate the same thing.
% We did not see such co-relation in our experiment also.
A limitation of \solx is that it considers only one test file and
generates only one block of code.  In real-world projects, test code
can be spread across multiple files or blocks of code.
% \solx  cannot generate tests that must be written across multiple files.
Despite that, \solx exceeded the state-of-the-art performance, so we
leave further improvements to future work.  We use the Python
\texttt{\small coverage} package for computing test coverage, but this
package can fail for various reasons, such as permission issues,
version incompatibility, or configuration problems. In \solx, we
computed coverage for all projects, including SymPy. However, upon
manual validation, we found the coverage information for SymPy to be
unreliable. Therefore, we removed coverage from the final
$\mathit{tddScore}$ metric for SymPy instances ($<\!\!15\%$ of the
total instances).  Note that coverage does not affect the reported
fail-to-pass scores.
% Given that coverage for fail-to-pass tests was consistently above 0.9 and fewer than 15\% of instances came from SymPy, this likely makes $<1.5$\% difference for the results.

%% file: conclusion.tex
\section{Conclusion}\label{sec:conclusion}

The primary contribution of this paper is \solx, a system for
generating tests from issue descriptions before issue resolution.
\solx outperforms the prior state of the art in fail-to-pass
tests generated while also costing less.  This paper also contributes
\tdd, a new benchmark for the same problem statement, mined from
real-world GitHub issues with strict filters and evaluation metrics.
Finally, this paper demonstrates that generated tests can improve
patches generated by SWE agents, helping them reach a precision of
between 65\% and 92\%. \tdd and \solx generated tests are at
{\small\url{https://github.com/IBM/TDD-Bench-Verified}}.